\documentclass{aa}  
\usepackage{graphicx}
\usepackage[varg]{txfonts}
\usepackage{xcolor}
\usepackage{booktabs,siunitx}

\newcommand{\juliet}{\texttt{juliet}}
\newcommand{\serval}{\texttt{serval}}
\newcommand{\george}{\texttt{george}}

\newcommand{\tess}{TESS}
\newcommand{\gaia}{\textit{Gaia}}

\newcommand{\au}{au}
\def\ms{m\,s$^{-1}$}

\newcommand{\addref}[1]{(add ref)}

\usepackage[normalem]{ulem} 
\newcommand\crossout{\bgroup\markoverwith{\textcolor{red}{\rule[0.5ex]{2pt}{1.2pt}}}\ULon} 

\usepackage[breaklinks, colorlinks, citecolor=blue, linkcolor=blue]{hyperref}
\newcommand{\abs}[1]{|#1|}

\defcitealias{Carleo2020_ADLeo_GIARPS}{C20}
\defcitealias{Robertson2020_ADLeo_HPF}{R20}
\defcitealias{Tuomi2018_ADLeo}{T18}

\begin{document} 

   \title{The CARMENES search for exoplanets around M dwarfs}
   \subtitle{
   Stable radial-velocity variations at the rotation period of AD~Leonis - A test case study of current limitations to treating stellar activity\thanks{Tables~\ref{tab:carmvisdata},~\ref{tab:carmnirdata},~\ref{tab:harpsdata},~\ref{tab:hiresdata}, and additional data (i.e., stellar activity indicators as shown in Figs.~\ref{fig:carmvisactivity},~\ref{fig:carmvniractivity}, and~\ref{fig:harpsactivity}) are available in electronic form at the CDS via anonymous ftp to \url{cdsarc.u-strasbg.fr} (130.79.128.5)}}
    \titlerunning{AD~Leo: stellar activity in radial velocity}

\author{
D.~Kossakowski\inst{1}
\and M.~K\"urster\inst{1} 
\and Th.~Henning\inst{1} 
\and T.~Trifonov\inst{1,2} 
\and J.\,A.~Caballero\inst{3} 
\and M.~Lafarga\inst{4,5,6} 
\and F.\,F.~Bauer\inst{7} 
\and S.~Stock\inst{8}
\and J.~Kemmer\inst{8}
\and S.\,V.~Jeffers\inst{9} 
\and P.\,J.~Amado\inst{7} 
\and M.~P\'erez-Torres\inst{7} 
\and V.\,J.\,S.~B\'ejar\inst{10,11} 
\and M.~Cort\'es-Contreras\inst{3} 
\and I.~Ribas\inst{5,6} 
\and A.~Reiners\inst{12} 
\and A.~Quirrenbach\inst{8} 
\and J.~Aceituno\inst{7} 
\and D.~Baroch\inst{5,6} 
\and C.~Cifuentes\inst{3} 
\and S.~Dreizler\inst{12} 
\and A.~Hatzes\inst{13} 
\and A.~Kaminski\inst{8}
\and D.~Montes\inst{14} 
\and J.\,C.~Morales\inst{5,6} 
\and A.~Pavlov\inst{1} 
\and L.~Pe\~na\inst{7} 
\and V.~Perdelwitz\inst{15,16} 
\and S.~Reffert\inst{8}
\and D.~Revilla\inst{7} 
\and C.~Rodr\'iguez~L\'opez\inst{7} 
\and A.~Rosich\inst{5,6} 
\and S.~Sadegi\inst{8,1}
\and J.~Sanz-Forcada\inst{3} 
\and P.~Sch\"ofer\inst{7,12} 
\and A.~Schweitzer\inst{16} 
\and M.~Zechmeister\inst{12}
}
\institute{
Max-Planck-Institut f\"{u}r Astronomie, K\"{o}nigstuhl  17, 69117 Heidelberg, Germany\\
\email{kossakowski@mpia.de}
\and Department
 of Astronomy, Sofia University ``St Kliment Ohridski'', 5 James Bourchier Blvd, BG-1164 Sofia, Bulgaria 
\and Centro de Astrobiolog\'ia (CSIC-INTA), ESAC, Camino bajo del castillo s/n, 28692 Villanueva de la Ca\~nada, Madrid, Spain
\and Department of Physics, University of Warwick, Gibbet Hill Road, Coventry CV4 7AL, United Kingdom
\and Institut de Ci\`encies de l'Espai (ICE, CSIC), Campus UAB, C/ de Can Magrans s/n, 08193 Cerdanyola del Vall\`es, Spain
\and Institut d'Estudis Espacials de Catalunya (IEEC), C/ Gran Capit\`a 2-4, 08034 Barcelona, Spain
\and Instituto de Astrof\'isica de Andaluc\'ia (CSIC), Glorieta de la Astronom\'ia s/n, 18008 Granada, Spain
\and Landessternwarte, Zentrum f\"ur Astronomie der Universit\"at Heidelberg, K\"onigstuhl 12, 69117 Heidelberg, Germany
\and Max-Planck-Institut f\"ur Sonnensystemforschung, Justus-von-Liebig Weg 3, 37077 G\"ottingen, Germany
\and Instituto de Astrof\'isica de Canarias (IAC), 38205 La Laguna, Tenerife, Spain
\and Departamento de Astrof\'isica, Universidad de La Laguna, 38206 La Laguna, Tenerife, Spain
\and Institut f\"ur Astrophysik, Georg-August-Universit\"at, Friedrich-Hund-Platz 1, 37077 G\"ottingen, Germany
\and Th\"uringer Landessternwarte Tautenburg, Sternwarte 5, 07778 Tautenburg, Germany
\and Departamento de F\'isica de la Tierra y Astrof\'isica \& IPARCOS-UCM (Instituto de F\'isica de Part\'iculas y del Cosmos de la UCM), Facultad de Ciencias F\'isicas, Universidad Complutense de Madrid, E-28040 Madrid, Spain
\and Department of Physics, Ariel University, Ariel 40700, Israel
\and Hamburger Sternwarte, Gojenbergsweg 112, 21029 Hamburg, Germany
}

\date{Received 15 April 2022 / Accepted 31 August 2022}

  \abstract
  {A challenge with radial-velocity (RV) data is disentangling the origin of signals either due to a planetary companion or to stellar activity. In fact, the existence of a planetary companion has been proposed, as well as contested, around the relatively bright, nearby M3.0\,V star AD~Leo at the same period as the stellar rotation of 2.23\,d.}
  {We further investigate the nature of this signal. We introduce new CARMENES optical and near-IR RV data and an analysis in combination with archival data taken by HIRES and HARPS, along with more recent data from HARPS-N, GIANO-B, and HPF. Additionally, we address the confusion concerning the binarity of AD~Leo.}
  {We consider possible correlations between the RVs and various stellar activity indicators accessible with CARMENES. We additionally applied models within a Bayesian framework to determine whether a Keplerian model, a red-noise quasi-periodic model using a Gaussian process, or a mixed model would explain the observed data best. We also exclusively focus on spectral lines potentially associated with stellar activity.}
  {The CARMENES RV data agree with the previously reported periodicity of 2.23\,d, correlate with some activity indicators, and exhibit chromaticity. 
  However, when considering the entire RV data set, we find that a mixed model composed of a stable and a variable component performs best. Moreover, when recomputing the RVs using only spectral lines insensitive to activity, there appears to be some residual power at the period of interest. We therefore conclude that it is not possible to determinedly prove that there is no planet orbiting in synchronization with the stellar rotation given our data, current tools, machinery, and knowledge of how stellar activity affects RVs. We do rule out planets more massive than 27\,M$_{\oplus}$ (= 0.084\,M$_\textnormal{Jup}$). Likewise, we exclude any binary companion around AD~Leo with M$\sin{i}$ greater than 3--6\,M$_\textnormal{Jup}$ on orbital periods $<14$\,yr.}{}


   \maketitle
%
\section{Introduction}

In the pursuit for Earth-like planets, modern spectrographs are pushing the limits by reaching \ms\ level precision or even tens of c\ms\ \citep[e.g., ESPRESSO;][]{espresso}, which is needed. However, the measurements begin to succumb to unwanted signals for planet searches. Intrinsic stellar variability in the form of dark spots, bright plages, and flares can produce radial-velocity (RV) variations that can conceal true planetary signals, or even masquerade as a fake planet which then can be effectively modeled with a Keplerian orbit. To help mitigate these stellar-activity-induced RV signals, a number of procedures are commonly put in place, such as different modeling approaches, smart data collection strategies, and extraction of particular spectral lines. 

Statistical techniques such as Gaussian process (GP) regression have been used by treating stellar activity behavior as a quasi-periodic signal \citep[e.g.,][]{Haywood2014,Rajpaul2015,Jones2017}. This approach may sometimes lead to better precision and accuracy of planetary parameters, especially those in the lower-mass regime \citep[e.g.,][]{Stock2020_YZCeti,Amado2021_gl393}. 
Likewise, these signals can be wavelength-dependent, usually with the amplitude decreasing in the redder regime of the spectrum, but still containing some residual effect depending on the star-spot configuration and temperature difference \citep[][and references therein]{Reiners2010}. 
For this reason, there is a push for higher-precision instruments covering the red and near-IR wavelength range such as CARMENES\footnote{Calar Alto high-Resolution search for M dwarfs with Exo-earths with Near-infrared and optical \'Echelle Spectrographs, \url{http://carmenes.caha.es}} \citep{CARMENES,CARMENES18}, GIARPS\footnote{GIAno \& haRPS-n, \url{https://www.tng.inaf.it/news/2017/04/04/giarps/}} \citep{Claudi2017_GIARPS}, HPF\footnote{The Habitable Zone Planet Finder, \url{https://hpf.psu.edu/}} \citep{Mahadevan2012,Mahadevan2014}, IRD\footnote{InfraRed Doppler instrument} \citep{Tamura2012_ird,Kotani2014_ird,Kotani2018_ird}, MAROON-X\footnote{Red-optical, high-resolution spectrograph with focus on mid- to late-type M dwarfs, \url{https://www.gemini.edu/instrumentation/maroon-x/}} \citep{Seifahrt2018_maroonx,Seifahrt2020_maroonx}, and SPIRou\footnote{A near-infrared spectropolarimeter/velocimeter, \url{https://spirou.omp.eu/}} \citep{Donati2020_spirou}.

Focusing on certain spectral lines as activity indicators sensitive to chromospheric (e.g., H$\alpha$, Ca~\textsc{ii}~infrared triplet) or photospheric (e.g., TiO) effects on active M dwarfs can oftentimes be successful in determining the star's rotational period \citep[see Fig. 11 in][]{Schoefer2019}. Even then, there is no universal approach because it is still not unique as to which activity indices do peak at the rotational period and under what conditions, as pointed out by \cite{Lafarga2021}.
Moreover, efforts for identifying which spectral lines, in general (i.e., not focusing on already-known specific lines), seem to be more activity-sensitive than others have been fruitful for a selection of G--K dwarfs \citep[e.g.,][]{Wise2018activelines,Dumusque2018_spectrallines,Lisogorskyi2019activityalphacenb,Ning2019linesbayes,Cretignier2020indivline,Thompson2020newactind}. Such an approach proves to be rather challenging for M dwarfs, where the spectra contain a forest of blended lines, making it almost impossible to find the continuum \citep{Merrill1962,Boeshaar1976,Kirkpatrick1991,AlonsoFloriano2015}. This approach, however, seems to have been successful for stars that show a clear stellar activity impact on the RVs (e.g., \object{EV~Lac}, Lafarga priv. comm.).

We turn our attention specifically to the active mid-type M dwarf AD~Leo, a star whose stellar rotation period of 2.23\,d presents itself both in photometry and RVs \citep{Morin2008,Tuomi2018_ADLeo,Carleo2020_ADLeo_GIARPS,Robertson2020_ADLeo_HPF}. 
Despite its strong flaring activity manifesting itself at many wavelengths \citep{Buccino2007,Rauer2011,Tofflemire2012,Vidotto2013}, AD~Leo has been included in a number of studies addressing the existence of planets orbiting this star. 
\cite{Tuomi2018_ADLeo}, referred to as \citetalias{Tuomi2018_ADLeo} hereinafter, first suggested that a planet may be orbiting AD~Leo in a 1:1 spin-orbit resonance since it proved to be difficult to simultaneously explain both the photometry and RV measurements using a variety of star-spot scenarios. Furthermore, they claimed that the RV measurements were time- and wavelength-independent, and the putative planet exhibited a semi-amplitude of $\sim$19\,\ms. Despite some evidence for solely stellar activity behavior, \citetalias{Tuomi2018_ADLeo} concluded that AD~Leo is an active M dwarf hosting a hot Jupiter ($\geq 0.2 M_\textnormal{Jup}$) in a 1:1 spin-orbit resonance. 

However, the existence of this hot Jupiter around AD~Leo was challenged. \citet[][referred to as \citetalias{Carleo2020_ADLeo_GIARPS} hereinafter]{Carleo2020_ADLeo_GIARPS} investigated the 2.23\,d signal and obtained observations with GIARPS at the 3.56\,m Telescopio Nazionale Galileo \citep{Claudi2017_GIARPS}. Even though the 2.23\,d signal was persistent, the amplitude heavily diminished as a function of wavelength and of time. Simultaneous photometric data from STELLA showed a shift of $\sim$0.25 in phase ($\sim$0.6\,d) in comparison to the HARPS-N RV curves. Therefore, \citetalias{Carleo2020_ADLeo_GIARPS} disputed the argument posed by \citetalias{Tuomi2018_ADLeo}, concluding that the RV modulation is not compatible with a planetary companion. Shortly after, the conclusions by \cite{Robertson2020_ADLeo_HPF}, referred to as \citetalias{Robertson2020_ADLeo_HPF} hereinafter, were also in line with \citetalias{Carleo2020_ADLeo_GIARPS}, as they likewise observed a decrease in amplitude between the two observing seasons using HPF spectroscopic data. At the time of submission, additional data from SPIRou and SOPHIE\footnote{Optical \'echelle spectrograph, \url{http://www.obs-hp.fr/guide/sophie/sophie-eng.shtml}} also indicated no evidence for a corotating planet (Carmona et al. in prep.).

This concept of such a hot Jupiter around an M dwarf in a synchronized rotation is unique. Many studies addressing the correlation between the orbital period ($P_\textnormal{orb}$) and the stellar rotation period ($P_\textnormal{rot}$) have thus far been focused on more solar-like transiting host stars rather than M dwarfs, using \textit{Kepler} data \citep{kepler}.
Findings from \cite{McQuillan2013} and \cite{WalkowiczBasri2013} suggest a clear absence of close-in planets ($P_\textnormal{orb} \lesssim$ 2–3 days) around rapidly rotating stars ($P_\textnormal{rot} \lesssim$ 5–10 days), where planets with shorter periods were nearly synchronous ($P_\textnormal{orb} \sim P_\textnormal{rot}$). 
\cite{Teitler2014} proposed that the cause was orbital decay of close-in planets by their host stars. Besides, with regards to M dwarfs, \cite{Newton2016} showed that typical stellar rotation periods would often match the periods of planets in their habitable zone. 
Regardless, there is a dearth of close-in massive planets around M dwarfs where only a handful of such are known, such as NGST-1~b \citep{Bayliss2018_ngts} or TOI-519~b \citep[$M \lesssim14\,\textnormal{M}_\textnormal{Jup}$;][]{Parviainen2021}. This is not a result of observational bias as these types of planets would be simple to detect through transits and precise RVs.
AD~Leo was even used as a test case to check the effects of XUV irradiation on planet atmospheres \citep{Chadney2016}, considering that the ionsphere can play an influential role in regulating the stability of upper atmospheres on giant planets.
Therefore, understanding the case of AD~Leo could aid in finding the key methods for other potential planet detections that suffer from ambiguities due to 1:1 spin-orbit resonances.

In our focused study using optical and near-IR RV measurements, we claim that one cannot prove (or disprove) the suggested planetary signal in the AD~Leo system, as the problem is completely degenerate and we are limited by our current state-of-the-art machinery and measurements. We organize the paper as follows. 
In Sect.~\ref{sec:adleo}, we first introduce the active M dwarf AD~Leo and investigate its presumed binarity for the first time.
All available RV data used for the analysis are presented in Sect.~\ref{sec:data}. 
We turn our focus on the chromaticity of the RV signal for the CARMENES data in Sect.~\ref{sec:wavelengthdependence}. Still concentrating on the CARMENES data, in Sect.~\ref{sec:spectrallines} we perform a spectral line analysis for the target. 
Then combining all available spectroscopic data for the first time, we explore various modeling techniques within the Bayesian framework in Sect.~\ref{sec:timedependence}.
Finally, in Sect.~\ref{sec:discussion}, we present a discussion on our results as well as suggestions to break the degeneracy of this situation. We summarize our conclusions in Sect.~\ref{sec:conclusions}.


\section{AD~Leo} \label{sec:adleo}

\subsection{Stellar parameters}

\begin{table}
\centering 
\small
\begin{center}
\caption{Stellar parameters of AD~Leo.}
\label{tab:stellarparams}
\centering
\begin{tabular}{lcr}
\hline \hline
\noalign{\smallskip}
Parameter & Value & Reference \\
\hline
    \noalign{\smallskip}
    \multicolumn{3}{c}{\textit{Identifiers}}\\
    \noalign{\smallskip}
BD      & +20 2465 & Arg1861 \\
Ci      & 18 1244   & Por1915 \\
GJ      & 388      & Gli1957 \\
Karmn   & J10196+198 & Cab2016 \\
    \noalign{\smallskip}
    \multicolumn{3}{c}{\it Astrometry and kinematics}\\
    \noalign{\smallskip}
$\alpha$ (epoch J2016.0) & 10:19:35.7 & \gaia\ EDR3 \\ 
$\delta$ (epoch J2016.0) & 19:52:11.3 & \gaia\ EDR3 \\ 
$\mu_\alpha \cos \delta$ (mas\,yr$^{-1}$)    & $-498.62 \pm 0.03$    & \gaia\ EDR3   \\
$\mu_\delta$ (mas\,yr$^{-1}$)                 & $-43.43 \pm 0.03$   & \gaia\ EDR3     \\
$\pi$ (mas)     & $201.41 \pm 0.03$     & \gaia\ EDR3 \\
$d$ (pc)        & $4.964 \pm 0.001 $    & \gaia\ EDR3 \\
$\gamma$ (k\ms) & $12.286 \pm 0.021$ & Laf2020 \\
$U$ (k\ms)  & $-14.929 \pm 0.010$ & This work \\ 
$V$ (k\ms) & $-7.444 \pm 0.007$ & This work \\ 
$W$ (k\ms) & $+3.391 \pm 0.017$ & This work \\ 
Galactic population & Young disk & Mon2001 \\ 
Stellar kinematic group & Castor & LS2010 \\
    \noalign{\smallskip}
    \multicolumn{3}{c}{\it Key photometry}\\
    \noalign{\smallskip}
$G$ (mag)    & $8.2041 \pm 0.0015$   & \gaia\ EDR3 \\
$J$ (mag)    & $5.449 \pm 0.027$        & 2MASS\\
    \noalign{\smallskip}
    \multicolumn{3}{c}{\textit{Photospheric parameters}}\\
    \noalign{\smallskip}
Spectral type   & M3.0\,V   & AF2015 \\ 
$T_\textnormal{eff}$ (K)    & $3477 \pm 23$ & Mar2021 \\
$\log{g}$ (cgs) & $5.12 \pm 0.12$ & Mar2021 \\
{[Fe/H]} (dex)  & $-0.19 \pm 0.12$ & Mar2021 \\
<Bf> (G) & $3357 \pm 172$ & Rein2022\\ 
    \noalign{\smallskip}
    \multicolumn{3}{c}{\textit{Activity}}\\
    \noalign{\smallskip}
$v \sin{i}$ (k\ms)  & $2.4 \pm 1.5$ & This work\tablefootmark{a} \\ 
$P_\textnormal{rot}$ (d)  & $2.2270^{+0.0010}_{-0.0011}$ & This work\tablefootmark{b} \\
pEW(H$\alpha$) (\AA) & $-3.73 \pm 0.41$ & Fuhr2022 \\
$\log{\textnormal{H}\alpha/L_\textnormal{bol}}$ & $-3.614 \pm 0.003$ & Sch2019 \\
$\log{R'_\textnormal{HK}}$ & $-3.97 \pm 0.05$ & This work \\
$\log{L_\textnormal{X}/L_\textnormal{bol}}$ & --3.3 & Fav2000 \\    \noalign{\smallskip}
    \multicolumn{3}{c}{\textit{Physical parameters}}\\
    \noalign{\smallskip}
$L_\star$ ($10^{-5}\,L_\odot$) & $2359 \pm 11$  & This work \\ 
$R_\star$ ($R_\odot$)            & $0.4233 \pm 0.0057$  & This work \\ 
$\textnormal{M}_\star$ ($\textnormal{M}_\odot$)            & $0.423 \pm 0.012$ & This work \\ 
$i$ (deg) &  $12.9^{+8.4}_{-8.1}$ & This work\\ 
\noalign{\smallskip}
\hline
\end{tabular}
\tablefoot{
\tablefoottext{a}{This value was individually produced for AD~Leo using a chosen order selection and is consistent, but supersedes previous measurement \citep[e.g.,][]{Marfil2021}.}
\tablefoottext{b}{See Sect.~\ref{sec:modelresults} for the $P_\textnormal{rot}$ determination and Table~\ref{tab:protvalues} for other rotational periods found in the literature.}
}

\tablebib{
    2MASS: \cite{Skrutskie2006}; 
    AF2015: \cite{AlonsoFloriano2015};
    Arg1861: \cite{Bonnersternverzeichnis1860_2}; 
    Cab2016: \cite{Caballero2016}; 
    Fav2000: \cite{Favata2000};
    Fuhr2022: \cite{Fuhrmeister2022};
    \gaia\ EDR3: \cite{GaiaEDR3}; 
    Gli1957: \cite{Gliese1957}; 
    Hoj2019: \cite{Hojjatpanah2019};
    Laf2020: \cite{Lafarga2020};
    LS2010: \cite{LopezSantiago2010};
    Mar2021: \cite{Marfil2021};
    Mon2001: \cite{Montes2001};
    Rein2022: \cite{Reiners2022};
    Pass2019: \cite{Passegger2019};
    Por1915: \cite{Porter1915};
    Sch2019: \cite{Schoefer2019}.
} 
\end{center}
\end{table}

AD~Leo (GJ~388), an M3.0\,V star at a distance of slightly less than 5\,pc and with $V \sim$ 9.5\,mag, is one of the closest and brightest M dwarfs. Already tabulated in the Bonner Sternverzeichnis by \citet{Bonnersternverzeichnis1860_2}, AD~Leo has been the subject of numerous investigations in the last century 
\citep[e.g.,][]{Abell1959,Engelkemeir1959,Lang1983,SaarLinsky1985,HawleyPettersen1991,Hawley2003,OstenBastian2008,HuntWalker2012}. 

In Table~\ref{tab:stellarparams}, we list the stellar properties of AD~Leo.
In particular, we tabulate equatorial coordinates, proper motions, and parallax from the \gaia\ Early Data Release 3 \citep[EDR3;][]{GaiaEDR3} and absolute RV from \citet[][uncorrected of gravitational redshift for consistency with the previous literature]{Lafarga2020}, from which we recompute Galactocentric space velocities as in \citet{CortesContreras2016}.
The spectral type of \citet{AlonsoFloriano2015} superseded previous determinations \citep[e.g.,][]{JohnsonMorgan1953,Bidelman1985,Stephenson1986,KeenanMcneil1989}, while the photosphere parameters of \citet{Passegger2019} match previous CARMENES publications and are similar, but not identical, to those of \citet{RojasAyala2012}, \citet{Lepine2013}, \citet{Gaidos2014}, or \citet{Mann2015}.
With the effective temperature of \citet{Passegger2019}, the bolometric luminosity of \citet{Cifuentes2020}, and the Stefan-Boltzmann law we derived the stellar radius and, with the radius-mass relation of \citet{Schweitzer2019}, the stellar mass.
For compiling the most precise parameters of the activity indicators, we used the \texttt{SVO Discovery Tool}\footnote{\url{http://sdc.cab.inta-csic.es/SVODiscoveryTool/}}.
The projected rotational velocity $v\sin{i}$ was computed by us exactly as in \citet{Reiners2018b}, but on the newest CARMENES template spectra (Sect.~\ref{sec:data}).

As first reported by \cite{Montes2001}, the Galactocentric space velocity of AD~Leo is consistent with it belonging to the Galactic young disk \citep{Leggett1992}. Later, \cite{LopezSantiago2010} and \cite{Klutsch2014} proposed AD~Leo as a candidate member of the Castor moving group, in agreement with our latest kinematic data. The age of the Castor moving group, of about 300--500\,Myr \citep{Barrado_Navascues1998, Mamajek2013}, is consistent with age determinations for AD~Leo by \citet{Shkolnik2009}, \citet{Brandt2014}, and \citet{Meshkat2017}.

Such a young age partly explains the flares frequently observed in AD~Leo. The star has been known to exhibit activity ever since the first observed optical flare event in 1949 \citep{Gordon_1949}, followed by many others \citep[e.g.,][]{Liller1952, MacConnell1968, Pettersen1984, CrespoChacon2006}\footnote{Of the 70 reports of the IAU Information Bulletin on Variable Stars citing AD~Leo, only 18 did not have the star name in the title.}. 
Frequent flaring activity was further observed during an extreme ultraviolet (EUV) 1-month monitoring of AD~Leo \citep{SanzForcada2002}.
Later on, emission in the X-ray and radio regimes has also been observed \citep{Gurzadyan1971,Robinson1976}, and which has shown bursting radio emission at GHz frequencies and below \citep{OstenBastian2008,Villadsen2019}.
\citet{Muheki2020} and \citet{Namekata2020} have presented the most recent analyses on high-resolution optical spectroscopy and X-ray observations of flares on AD~Leo.

The young age of AD~Leo also explains the moderately large rotational velocity and short rotational period, of 2.23\,d, as well as X-ray, Ca~\textsc{ii} H\&K, and H$\alpha$ emission (see references in Table~\ref{tab:stellarparams}).
In addition, this star presents large RV variations: it shows a standard deviation larger than 20\,\ms, ($\sim 1.5 \times$ median absolute deviation RV as in \citealt{Tal-Or2019_HIRES} and \citealt{Grandjean2020}), which can be connected to stellar activity, the presence of a planetary companion, or both. 
For the derivation of $\log{R'_\textnormal{HK}}$, we used the method described by \cite{Volker2021}, albeit with a slight modification of the $k1$ passband used for normalization. A total of 316 archival spectra from ESPaDOnS (232 spectra), HARPS (58 spectra), HIRES (24 spectra) and NARVAL (2 spectra) were analyzed, yielding a mean value of $\log{R'_\textnormal{HK}}=-3.97\pm0.05$, which is in agreement with the result published by \cite{AstudilloDefru2017} based on the HARPS spectra alone.
We compute the stellar inclination, $i$, to be $\sim 13^\circ$ from $10^6$ MCMC realizations using $v \sin{i}$, $R$, and $P_\textnormal{rot}$ as provided in the stellar parameters table. Previous papers quoted values of $\sim20^\circ$ \citep{Morin2008} and $\sim15^\circ$ \citep{Tuomi2018_ADLeo}. 

%
%

\subsection{Photometric rotational period} \label{sec:phot}
\citetalias{Tuomi2018_ADLeo} took a closer look at the available photometry from ASAS-North and ASAS-South, as well as from short-cadence \textit{MOST} observations (see \citetalias{Tuomi2018_ADLeo} for photometry references).
To summarize, the ASAS photometry shows a long-term periodicity of around 4070\,d, most likely due to a stellar activity cycle. During the brightness minimum of this cycle, the signal at the rotation period of 2.23\,d seems to disappear turning up only during the brightness maximum, in which the \textit{MOST} observations were taken. These short-cadence \textit{MOST} data, taken over the course of 9\,d \citep{HuntWalker2012}, showed fluctuations with periodicity 2.23\,d, but also demonstrated behavior of slight phase shifts and amplitude and period variations, indicating that the stellar surface should be experiencing rapid evolution. Other observations taken with different instruments are in agreement with the 2.23\,d period (see Table~\ref{tab:protvalues}).

AD~Leo was additionally observed in three of the \tess\ sectors\footnote{\tess\ Web Viewing Tool: \url{https://heasarc.gsfc.nasa.gov/cgi-bin/tess/webtess/wtv.py} Accessed 6 April 2022.}, particularly in sector 45 (camera \#3 CCD \#4; 6 November 2021 to 02 December 2021), sector 46 (camera \#2 CCD \#2; 02 December 2021 to 30 December 2021), and in sector 48 (camera \#1 CCD \#4; 28 January 2022 to 26 February 2022). All sectors constitute short cadence 20-second and 2-minute integrations. Unfortunately, the pixel column was saturated for sectors 45 and 46 leading to unusable data. The sector 48 data were still valuable for determining a periodic variation. In doing so, we removed strong flare events (up to 275\,ppt) and any outliers on the 20-second integration data which were binned every 10 points for computational reasons. After applying a sinusoid model, the resulting amplitude and $P_\textnormal{rot}$ are $\sim$800\,ppm and $P_\textnormal{rot}$ of $2.2304 \pm 0.0014$\,d, respectively. 


\subsection{Hypothetical binary} \label{sec:binarity}

There is some confusion in the literature regarding the hypothetical multiplicity of AD~Leo.
First of all, in spite of being listed as ``$\gamma$~Leo~C'' in some astronomical databases (e.g., SIMBAD), AD~Leo is not a wide ($\rho \sim$ 5.7\,arcmin) physical companion to \object{Algieba}, which is a binary system of two G- and K-type giant stars visible to the naked-eye located eight times further \citep[\object{$\gamma^{01}$~Leo} + \object{$\gamma^{02}$~Leo};][]{vanLeeuwen2007,GaiaEDR3}.

Next, the Washington Double Star \cite[WDS;][]{WDS2001} catalog tabulates AD~Leo as a close binary candidate.
The presence of a hypothetical companion around AD~Leo was first inferred by \cite{Reuyl1943} from measurements of photographic plate images. 
He suggested an eccentric ($e=0.6$) and close-in ($a\sim 0.54$\,\au, projected angular separation of $\rho\sim 0.11$\,arcsec) orbit with an orbital period of 26.5\,years, making the companion a brown dwarf ($\textnormal{M} \approx 0.032 \textnormal{M}_\odot$). 
Later, \cite{vandeKamp1949} found indications of an astrometric trend from photographic plates that did not fit those orbital elements. 
Following up two decades later, \cite{Lippincott1969} found an ambiguous deviation from linear proper motion, 
thus deciding that it was inconclusive to determine whether there could be a variable proper motion due to a companion.


With speckle imaging at 7800\,\AA\ at the 3.6\,m Canada-France-Hawai'i Telescope, \cite{Balega1984} resolved a companion candidate to AD~Leo at $\rho$ = 0.078\,$\pm$\,0.010\,arcsec ($r \approx$ 0.39\,\au) and position angle of $\theta$ = 39\,$\pm$\,4\,deg in 1981, which received the WDS name 10200+1950 and discoverer code BAG~32. 
Two additional measurements in 1983 were provided by \cite{BalegaBalega1985} at $\rho \sim$ 0.11\,arcsec, but with $\theta \sim$ 330\,deg. Afterwards, with lucky imaging in the $I$ band at the 1.5\,m Telescopio Carlos S\'anchez, \cite{CortesContreras2017} resolved a candidate source, about 2.0\,mag fainter than AD~Leo, at $\rho$ = 0.195\,$\pm$\,0.061\,arcsec ($r \approx$ 0.97\,\au) and $\theta$ = 23.8\,$\pm$\,3.7\,deg in 2012.
However, it fell on the first Airy disk, and they were not able to detect it again in observations in 2015 with the same instrument setup.
From $\Delta I \approx 2.0$\,mag and the projected physical separation of about 0.97\,\au, \cite{CortesContreras2017} estimated an M5\,V spectral type for the companion candidate and an orbital period of $P \sim$ 1.5\,yr, which might be the source discovered by \cite{Balega1984}.
Because of this hypothetical companion, AD~Leo was initially discarded from the CARMENES guaranteed time observation target list \citep{Caballero2016,Reiners2018b}.

\begin{figure}
    \centering
    \includegraphics{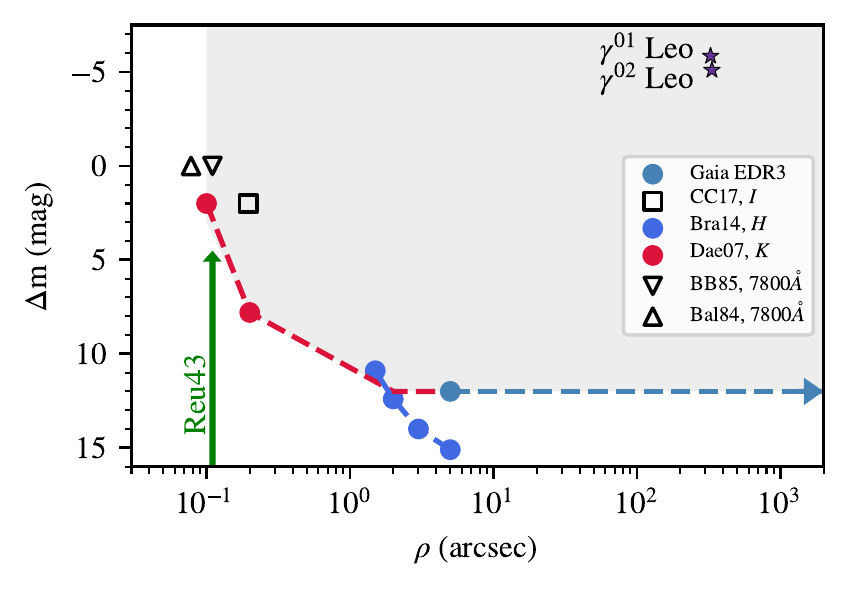}
    \caption{Magnitude difference versus projected angular separation diagram for observations of potential companions around AD~Leo. Detections should be above and to the right of the lines (gray shaded region). Inaccurate detection claims are indicated with open symbols. Measurements from \gaia\ EDR3 are indicated with an extended arrow to the right to demonstrate it continues for many more arcseconds. The upwards green arrow represents a rough difference in magnitude interval from the astrometric masses indicated by \cite{Reuyl1943}, which are difficult to pinpoint as they are subject to huge uncertainties. The bright binary $\gamma$~Leo in the background of AD~Leo is discussed in Sect.~\ref{sec:binarity}. References --
    CC17: \cite{CortesContreras2017}; Bra14: \cite{Brandt2014}; Dae07: \cite{Daemgen2007}; BB85: \cite{BalegaBalega1985}; Bal84: \cite{Balega1984}; Reu43: \cite{Reuyl1943}, astrometry; \gaia\ EDR3: \cite{GaiaEDR3}, astrometry. 
    }
    \label{fig:magseparation}
\end{figure}

Further attempts at resolving the companion proved to be unsuccessful, such as 
additional speckle imaging by \cite{Docobo2006}. 
With adaptive optics at 8\,m-class telescopes, \citet{Daemgen2007} and \citet{Brandt2014} imposed very strict upper limits to the presence of companions, of $\Delta K_s \sim$ 7.8\,mag at $\rho \ge$ 0.5\,arcsec and $\Delta H \sim$ 10.9\,mag at $\rho \ge$ 1.5\,arcsec, respectively.
From Fig.~7 in \citet{Daemgen2007}, their Altair/Gemini North observations discarded any companion of $\Delta K_s \sim$ 2.0\,mag at $\rho \ge$ 0.1\,arcsec.
No trace of companions is seen either in archival images obtained in 2005 with the \textit{Hubble Space Telescope} (ACS/HRC in F330W) and in 2018 with the Very Large Telescope (SPHERE in $H$)\footnote{Near-infrared archival images obtained in 1999 with {\it Hubble} NICMOS/NIC2 and in 2009 with VLT NAOS+CONICA saturate and are not useful at very close separations.}.
Therefore, the findings of \cite{Balega1984}, \cite{BalegaBalega1985}, and \cite{CortesContreras2017} are most probably not related to a real physical companion, unless the orbital motion can explain the later nondetections.
Furthermore, the \gaia\ EDR3 renomalized unit weight error (\texttt{RUWE}) for AD~Leo is 1.15, below the critical threshold of 1.40, the along-scan observations (\texttt{astrometric\_gof\_al}) is 3.22, and the excess noise of the source (\texttt{astrometric\_excess\_noise}) is 0.16, all of which indicate that it is most likely a single source \citep{Lindegren2018}. To rule out any wide companions, we searched for objects with common parallax and proper motions up to a projected physical separation of 100\,000\,\au, as in \cite{Caballero2022}, and found no hints of any wide potential companion.
Figure~\ref{fig:magseparation} illustrates all the mentioned findings from the literature.
The binary issue was not addressed by \citetalias{Tuomi2018_ADLeo}, \citetalias{Carleo2020_ADLeo_GIARPS}, or \citetalias{Robertson2020_ADLeo_HPF}.
The absence of any close companion to AD~Leo is further investigated with our spectroscopic data in Sect.~\ref{sec:adleoisasinglestar}.

\subsection{Evidence for single star in spectroscopic data} \label{sec:adleoisasinglestar}
As a first approach to search to address the potential stellar binarity and look for traces of it, we combined all the RV data (presented in Sect.~\ref{sec:data}) to look for long-term trends. However, this method presents its challenges as there is no temporal overlap between the older and newer data sets and each instrument has its own, unknown zero-point offset. We nonetheless attempted a grid search for a Keplerian signal for which we stepped through the period, amplitude, eccentricity and argument of periastron parameter space, where each instrument also had its own offset and jitter term\footnote{Jitter terms are added in quadrature to the given error bars for each respective instrument.} \citep[e.g.,][]{Baluev2009}. One then adapts the values with the best log likelihood. Unfortunately, the lack of temporal overlap led to strong ambiguities and degeneracies. We therefore did not find any conclusive periodicity or indicative linear trend in this RV analysis. 

We continued to investigate whether there is evidence for the presence of a companion within our CARMENES spectra that could affect the analysis presented later in this work. We computed a 1D cross-correlation function (CCF) with a binary mask over a large RV range and did not find any hint of a clear secondary peak, neither in the VIS nor the NIR data. To be certain, we also ran \texttt{todcor} \citep{Zucker1994_TODCOR}, which computes a 2D CCF to get the RVs of the two components simultaneously, and the results showed no evidence for a companion as well. The 2D CCF method with \texttt{todcor} is well suited in the case of double-lined binaries, but since the secondary signal seems to be either too weak if expected to be $\sim$10\,k\ms\ away from the primary, or too hidden if expected to be very close to the primary, we concluded that no secondary heavily distorts the CCF profile and, therefore, would not cause any noticeable effects on the CCF parameter values. 
Therefore, we first discard the presence of a nearly equal brightness double-lined spectroscopic binary from the \texttt{todcor} analysis. Secondly, based on our RV time series analysis, we rule out companions with minimum masses greater than 3--6\,$\textnormal{M}_\textnormal{Jup}$ using orbital periods of less than $\sim$2--14\,yr and amplitudes 3--5 greater than the rms of the earlier data set instruments (i.e., HARPS, HIRES), while assuming $\sin{i}\neq0$.
We therefore conclude that there is no evidence for a stellar binary or brown dwarf companion of AD~Leo within this parameter regime.

\subsection{Bursting radio emission from the AD~Leo system} \label{sec:adleoradio}

Radio bursts from AD~Leo have been known for quite a long time, and have been ascribed to (coherent) plasma emission \citep{Stepanov2001, OstenBastian2008}. More recently, \citet{Villadsen2019} detected both short-duration (seconds to minutes) and long-duration (30 minutes or more) bursts of AD~Leo, using ultrawide-band VLA observations (in the ranges 0.2--0.5\,GHz and continuous frequency coverage between 1 and 6\,GHz) in several epochs between 2013 and 2015. 
The timescale of the short-duration events was consistent with the duration of ``space-weather'' events, such as the impulsive phase of a flare or a coronal mass ejection (CME) crossing the corona. The long-duration bursts, which lasted hours, and possibly extended for up to days or even longer times, between observing epochs (similarly to the case of \object{Proxima Centauri}; see, e.g., \citealt{PerezTorres2021} and references therein) require an ongoing electron acceleration mechanism during the burst. Candidate acceleration processes are found within the paradigms of solar radio bursts and periodic radio aurorae produced by ultracool dwarfs and planets.

The emission mechanism responsible for the aurorae from stars and planets alike  is the electron cyclotron maser (ECM) instability \citep{Melrose1982}, whereby plasma processes within the star (or planet) magnetosphere generate a population of unstable electrons that amplifies the  emission.
The characteristic frequency of the ECM emission is given by the electron gyrofrequency, $\nu_G = 2.8\,B$\,MHz, where $B$ is the local magnetic field in the source region, in Gauss. ECM emission is a coherent mechanism that yields broadband ($\Delta\,\nu \sim \nu_G/2$), highly polarized (sometimes reaching 100\%), amplified nonthermal radiation. The long-duration bursts seen for AD~Leo by \citep{Villadsen2019} between 1 and 6\,GHz are characteristic of ECM emission, implying electron densities $\lesssim 10^9$ cm$^{-3}$, and originate in regions with field strengths of 0.36--2.1\,kG (in fundamental emission) or 0.18--1.1\,kG (if second harmonic). Since the surface magnetic field of AD~Leo is estimated to be of $\simeq 3.36$\,kG \citep{Reiners2022}, the above field strengths are likely to occur relatively close to the star, at a height less than about one stellar radius.

Auroral cyclotron maser emission is powered by the acceleration of confined electrons with energies between $\sim$10\,keV and up to 1\,MeV energies. 
For substellar objects, the currents could be driven by the breakdown of rigid corotation of magnetospheric plasma with the object's magnetic field, for example, due to the interaction between a rotating magnetosphere and the interstellar medium. Corotation breakdown observed in a number of ultracool (UCD) dwarfs with periods less than about 3\,hr seem to be rotation powered, with radio powers of up to a few times $10^{13}$\,erg\,s$^{-1}$\,Hz$^{-1}$ \citep[e.g.,][]{Turnpenney2017}. AD~Leo has a rotation period of $\sim$54\,hr. Assuming coronal parameters similar to radio-loud UCDs, any corotation breakdown in AD~Leo will generate a radio power that is $\sim 6\times 10^{11}$\,erg\,s$^{-1}$\,Hz$^{-1}$, or less than about 100 $\mu$\,Jy. Therefore, corotation breakdown, while is likely to contribute to the overall radio power observed from AD~Leo, cannot account for the observed radio bursts, which reach peaks of tens of mJy at GHz frequencies \citep{Villadsen2019}. A discussion of further potential mechanisms, such as star-planet interaction, to explain the measurements is continued in Sect.~\ref{sec:starplanetinteractoin}.

\section{Spectroscopic data} \label{sec:data}

\begin{figure*}
\centering
\includegraphics[width=\hsize]{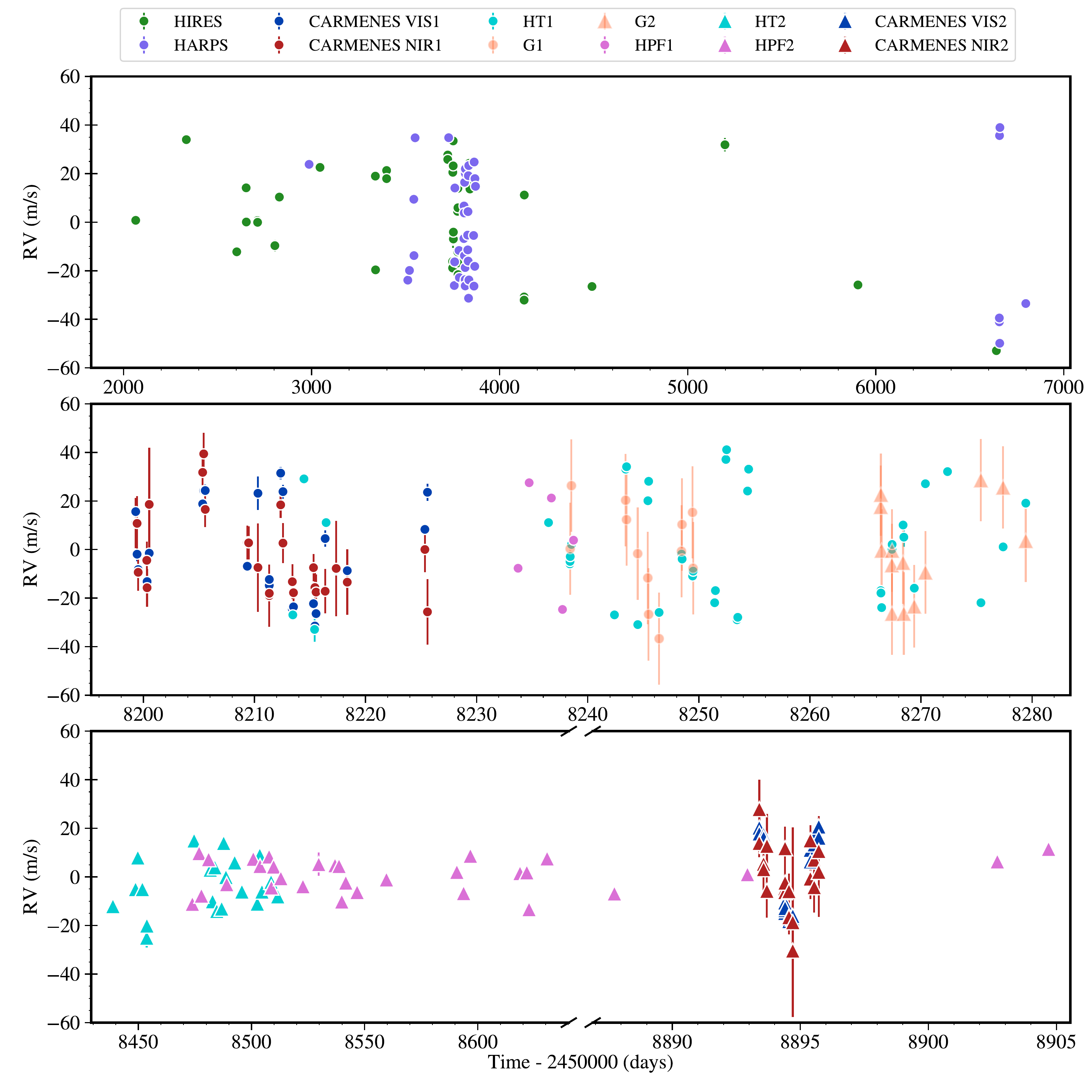}
\caption{Time series of all RVs with instrumental offsets accounted for. The RV uncertainties are included, though many are too small to be seen in the plots. The HIRES and HARPS data span a large time range and the rest of the data come in $\sim$12 years later after the time when the majority of the previous data were taken. The time axis is interrupted in several places, and stretched differently between the individual sections. Additionally, the majority of HIRES and HARPS time series are overlapping each other. Four of the HARPS-N data overlap with CARMENES data; and the GIANO-B data are taken all within the first observing run for HARPS-N. The first season of HPF data overlaps with the HARPS-N and GIANO-B data sets whereas the HPF second season overlaps with the HARPS-N and CARMENES second season.}
\label{fig:allrv_timeseries}
\end{figure*}

\begin{figure}
\centering
\includegraphics[width=\hsize]{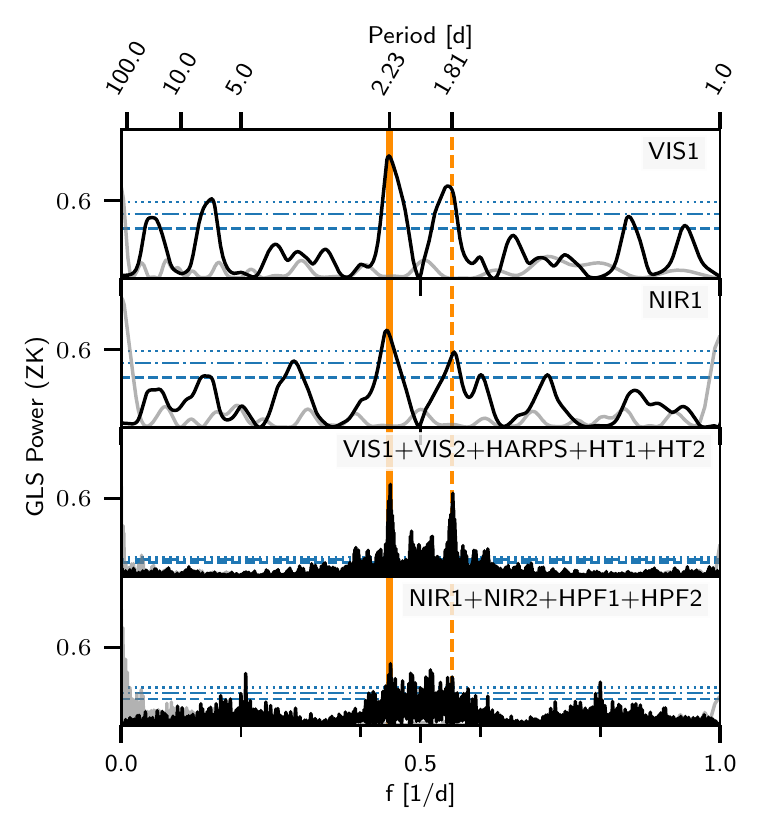}
\caption{GLS periodograms (black) and window functions (gray) for the CARMENES VIS1 and NIR1 RV data sets, as well as for the combined optical (i.e., VIS1, VIS2, HARPS, HT1, and HT2) and near-infrared (i.e., NIR1, NIR2, HPF1, HPF2) instruments, with offsets accounted for. The horizontal lines correspond to FAP levels of 10\% (dashed), 1\% (dash-dotted), and 0.1\% (dotted). These are computed by using the randomization technique for 10\,000 samples. The orange solid line corresponds to the rotational period of 2.23\,d, whereas the orange dashed line represents the 1.81\,d alias signal due to daily sampling.}
\label{fig:periodogramrv}
\end{figure}

The Doppler data that are used in our analysis are described below. 
All data and spectrograph descriptions are summarized in Table~\ref{tab:rvinstruments}. Data used for the analysis are displayed in Fig.~\ref{fig:allrv_timeseries}, along with the generalized Lomb-Scargle (GLS) periodograms \citep{Zechmeister2009_GLS} of the data in Fig.~\ref{fig:periodogramrv}. 

\begin{table*}
\caption{Summary of the extensive spectroscopic data set for AD~Leo. All instruments used for the analysis along with their wavelength range, spectral resolution, and number of spectra are listed.}
\label{tab:rvinstruments}
\centering
\scriptsize
\begin{tabular}{l c c c c c c c c c c c c}
\hline\hline
\noalign{\smallskip}
Instrument & \multicolumn{2}{c}{CARMENES VIS} & \multicolumn{2}{c}{CARMENES NIR} & HIRES & HARPS & \multicolumn{2}{c}{HARPS-N} & \multicolumn{2}{c}{GIANO-B} & \multicolumn{2}{c}{HPF} \\
\noalign{\smallskip}
\hline
\noalign{\smallskip}
$\Delta \lambda$ (nm) & \multicolumn{2}{c}{520--960} &\multicolumn{2}{c}{960--1710} & 500-620 & 380--690 & \multicolumn{2}{c}{390--680} & \multicolumn{2}{c}{970--2450} & \multicolumn{2}{c}{810--1280}\\
$R$ & \multicolumn{2}{c}{94\,600} & \multicolumn{2}{c}{80\,400} & 60\,000 & 115\,000 & \multicolumn{2}{c}{115\,000} & \multicolumn{2}{c}{50\,000} & \multicolumn{2}{c}{55\,000}\\
\noalign{\smallskip}
Subset \tablefootmark{a} & VIS1 & VIS2 & NIR1 & NIR2 & & & HT1 & HT2 & G1 & G2 & HPF1 & HPF2 \\
\cmidrule(l{1.5em}r{1.5em}){2-2}
\cmidrule(l{1em}r{1em}){3-3}
\cmidrule(l{1.5em}r{1.5em}){4-4}
\cmidrule(l{1em}r{1em}){5-5}
\cmidrule(l{1.5em}r{1.5em}){8-8}
\cmidrule(l{1.5em}r{1.5em}){9-9}
\cmidrule(l{1.75em}r{1.75em}){10-10}
\cmidrule(l{1.75em}r{1.75em}){11-11}
\cmidrule(l{1em}r{1em}){12-12}
\cmidrule(l{1em}r{1em}){13-13}
\# of spectra & 26 & 20 & 26 & 20 & 43 & 47 & 42 & 21 & 12 & 13 & 5 & 30\\
rms (\ms) & 18.54 & 15.63 & 16.44 & 13.61 & 20.74 & 24.23 & 21.97 & 10.51 & 17.92 & 18.41 & 19.01 & 6.68 \\
Start date & Mar 2018 & Feb 2020 & Mar 2018 & Feb 2020 & Jun 2001 & Dec 2003 & Apr 2018 & Nov 2018 & Apr 2018 & May 2008 & Apr 2018 & Dec 2018 \\
End date & Apr 2018 & Feb 2020 & Apr 2018 & Feb 2020 & Dec 2013 & May 2014 & Jun 2018 & Jan 2019 & May 2018 & Jun 2018 & Apr 2018 & Feb 2020 \\
\noalign{\smallskip}
\hline
\end{tabular}
\tablefoot{
\tablefoottext{a}{A subset is treated effectively as an independent data set.}
}
\end{table*}

\subsection{CARMENES} \label{sec:carmenes}

\begin{figure*}
\centering
\includegraphics[width=\hsize]{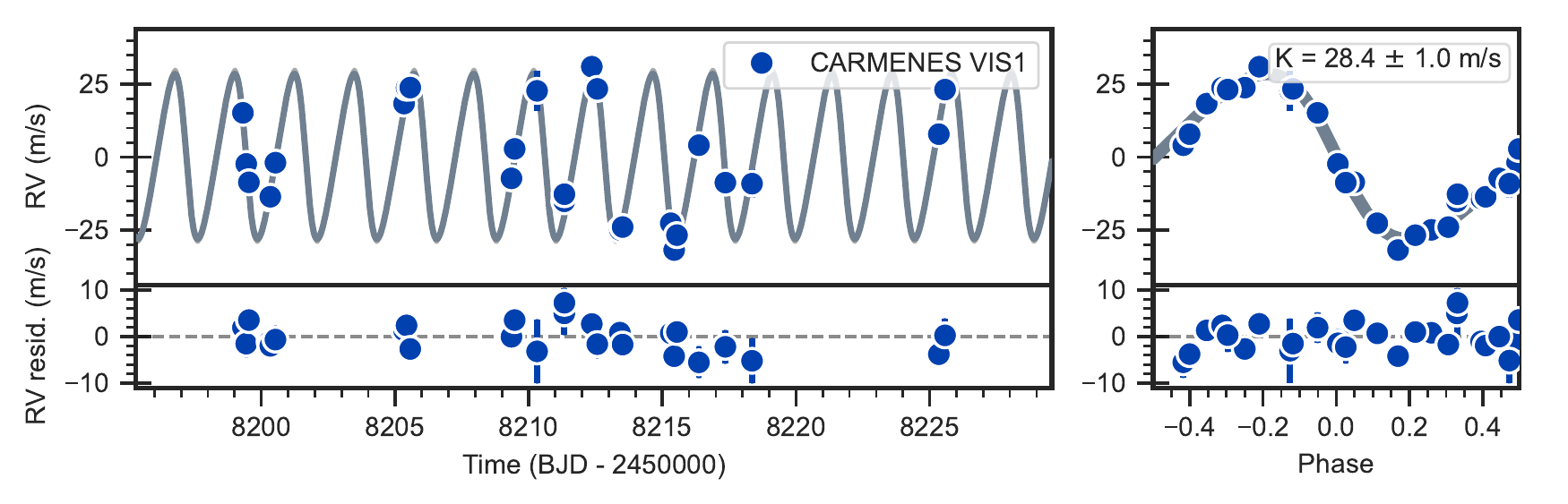}
\includegraphics[width=\hsize]{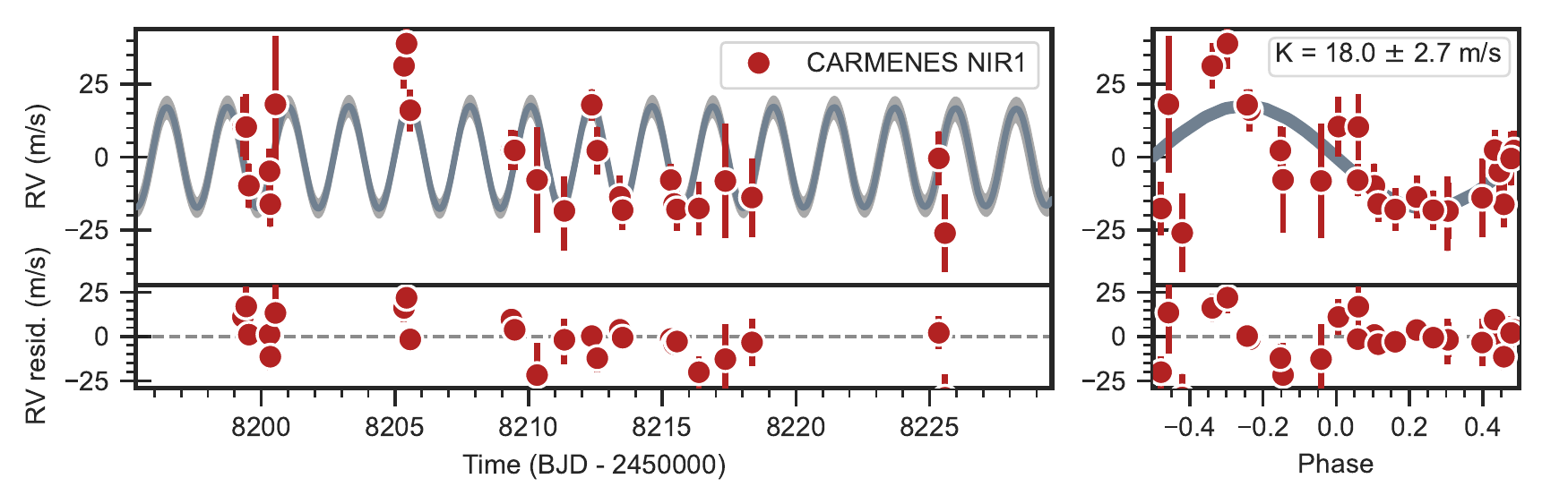}
\includegraphics[width=\hsize]{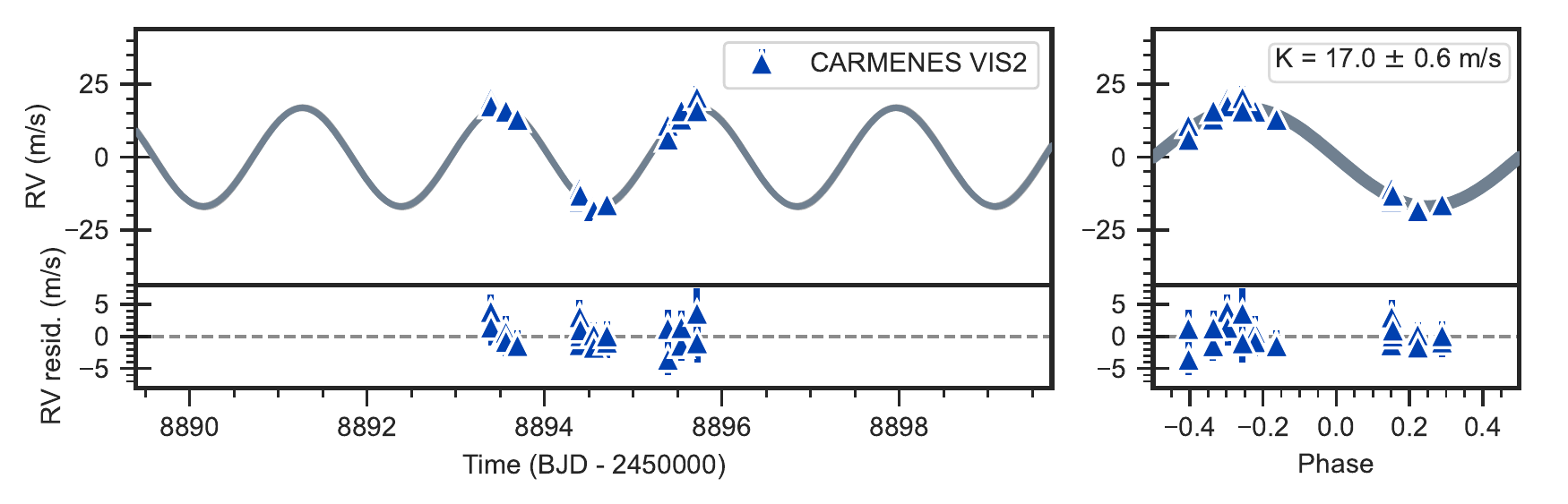}
\includegraphics[width=\hsize]{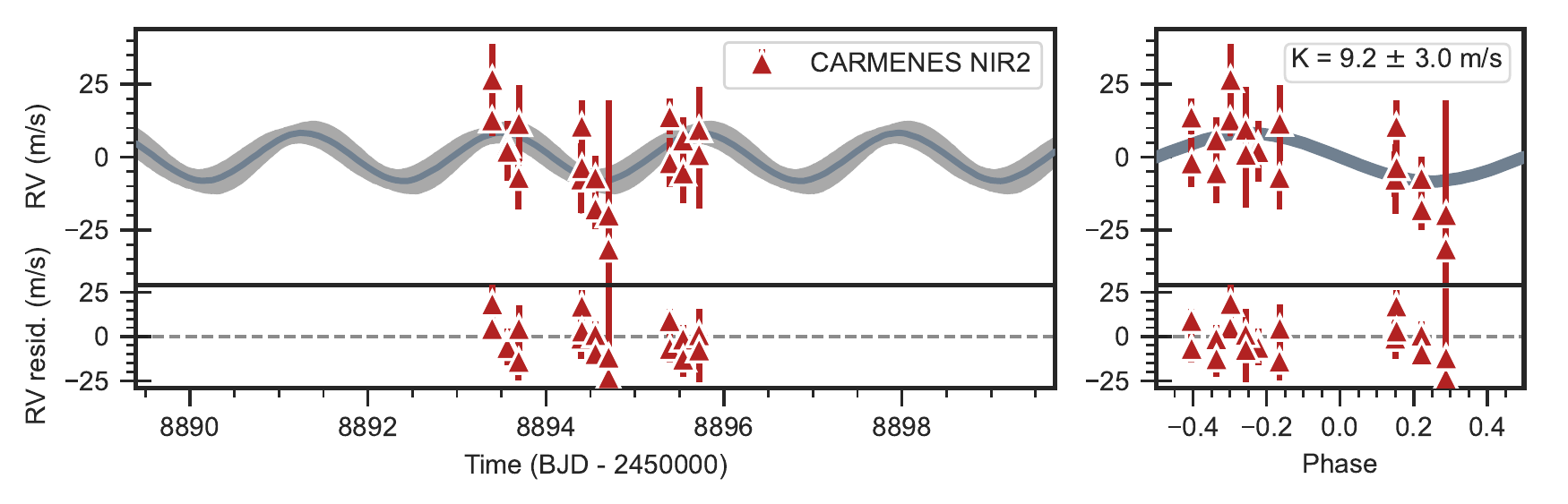}
\caption{Radial-velocity times series and phase-folded plots for the CARMENES VIS1, NIR1, VIS2, and NIR2 data (top to bottom) with residuals from the best-fit model. In some cases, especially in the panels with the VIS1 and VIS2 data, the uncertainties are too small to be seen in the plots. The best-fit models are overplotted (gray line) along with the 68\%\ posterior bands gray shaded regions).}
\label{fig:carmrv}
\end{figure*}

The CARMENES instrument is located at the 3.5\,m telescope at the Calar Alto Observatory in Spain. It is a dual arm instrument to produce RV measurements at optical (VIS) and near-infrared (NIR) wavelengths \citep[][see Table~\ref{tab:rvinstruments} for the instrument specifications]{CARMENES, CARMENES18}. 
We obtained 26 spectra for AD~Leo (Karmn J10196+198) over the time span of 25 days in 2018, notably 12 years after the majority of the HIRES and HARPS data were taken. On some nights, we observed AD~Leo multiple times to better sample the short period of 2.23\,d. We call this subset of data ``VIS1'' and ``NIR1'' given that AD~Leo was observed again later in 2020 as part of a DDT\footnote{Director's Discretionary Time} program, for which 20 data points were obtained over $\sim$3\,d, which we call subset ``VIS2'' and ``NIR2''. The RVs and several activity indicators (e.g., CRX -- chromatic index, dLW -- differential line width) from both subsets were first extracted with \serval\ \citep{Zechmeister2018_SERVAL}, so that they are corrected for barycentric motion, secular acceleration, and instrumental drift. The final RVs we use had nightly zero-points applied \citep{Tal-Or2019_HIRES,Trifonov2020}. Figure~\ref{fig:carmrv} displays the CARMENES data and the best respective models (detailed in Sect.~\ref{sec:timedependence}).
For calculating the CCF parameters within CARMENES, weighted binary masks, which depend on spectral type and $v \sin{i}$, are produced by coadding spectra corrected for tellurics \& RV shifts and then selecting pronounced minima \citep{Lafarga2020}. From there, the CCF parameters, namely, the bisector span (BIS), contrast, and full-width half-maximum (FWHM), are obtained for both the VIS and NIR channel.

\subsection{Archival RV data}
\paragraph{HARPS-South.} \label{sec:harps}
Spectroscopic data for AD~Leo are in the public archive from the High-Accuracy Radial velocity Planet
Searcher \citep[HARPS,][]{HARPS}, a high-resolution \'echelle spectrograph located at the ESO 3.6\,m telescope at La Silla Observatory in Chile. We retrieved a total of 52 spectra that were taken over a span of $\sim$4500\,d. Due to a fiber-upgrade intervention \citep{LoCurto2015}, we considered the HARPS data as coming from two separate instruments, before and after this upgrade. After the intervention, only five data points were taken close to one another (within 2\,d) and much later than the earlier data. These data also have a very low root-mean-square (RMS) deviation from their mean. We did not consider them for the further analysis since they could fit anywhere in a model with a large offset and therefore do not provide new insight. Thus, for the rest of the analysis, we used the 47 HARPS spectra from before the intervention. Additionally, the majority of data (i.e., 33 points) were taken over the course of $\sim$115\,d (January--May 2006). The spectra were first processed using \serval\ \citep{Zechmeister2018_SERVAL} to obtain the RVs and helpful stellar activity indicators \citep[e.g., CRX, dLW, H$\alpha$, see][for more explanation]{Zechmeister2018_SERVAL}. Nightly zero point corrections provided by \cite{Trifonov2020} were applied to the RVs\footnote{\url{https://www2.mpia-hd.mpg.de/homes/trifonov/HARPS_RVBank.html}} and are used in our analysis (Table~\ref{tab:harpsdata}).

\paragraph{HIRES.} \label{sec:hires}
We collected 43 spectra from the high-resolution spectrograph HIRES \citep{HIRES} mounted on the 10\,m Keck-I telescope located at the Mauna Kea Observatory in Hawai'i, which has been in service since 1994. The data were taken over the course of 4500 days. A majority of them (i.e., 22 points) were taken in 2005/2006 over 120\,d and, moreover, overlap with the higher cadence HARPS data (see Fig.~\ref{fig:allrv_timeseries}). \citet{Butler2017_HIRES} released a large RV database of 64\,480 observations for a sample of 1699 stars, which was later reanalyzed by \citet{Tal-Or2019_HIRES} for minor, though significant systematic effects, such as an RV offset due to the CCD upgrade in 2004, long-term drifts, and slight intra-night drifts. Therefore, we continued with the corrected HIRES RVs provided by \cite{Tal-Or2019_HIRES}, presented in Table~\ref{tab:hiresdata}. Data from HIRES are also specifically addressed further under Sect.~\ref{sec:otheractivityindicators}.

\subsection{Additional RV data from the literature} 
Radial velocities are taken directly from \citetalias{Carleo2020_ADLeo_GIARPS} and \citetalias{Robertson2020_ADLeo_HPF}, as described in the following paragraphs, and are used for the analysis as well.

\paragraph{GIARPS.} \label{sec:giarps}
We include spectroscopic data taken in GIARPS mode \citep{Claudi2017_GIARPS}, where high resolution spectroscopic measurements are obtained simultaneously with HARPS-N \citep[][extracted with the TERRA pipeline]{Cosentino2012_HARPSN} and with GIANO-B \citep[][]{Oliva2006_GIANO} reduced with the on-line DRS pipeline and the off-line \texttt{GOFIO} pipeline. Both instruments are located at the 3.58\,m Telescopio Nazionale Galileo (TNG) at the Roque de los Muchachos Observatory in La Palma, Spain. The GIARPS mode is similar to CARMENES in the sense it also addresses potential variations of a signal amplitude over a wide wavelength range. There were two runs of HARPS-N, with only the first run having simultaneous GIANO-B data; four of the HARPS-N data overlap with the CARMENES data. To stay consistent with \citetalias{Carleo2020_ADLeo_GIARPS}, we also consider two separate runs for HARPS-N and GIANO-B, and designate them as ``HT1'', ``HT2'', ``G1'', and ``G2'' for our analysis. Due to the high uncertainty on the data points from the GIANO-B instrument, we do not include them for our analysis.

\paragraph{HPF.} \label{sec:hpf}
Our last data set comes from the Habitable-zone Planet Finder (HPF), a stable NIR Doppler spectrograph that is designed to reach 1--3\,\ms\ RV precision for M dwarfs with the help of wavelength calibration via a custom NIR laser frequency comb \citep{Mahadevan2012,Mahadevan2014}. The spectrograph is installed at the 10\,m Hobby-Eberly Telescope at McDonald Observatory in Texas. The HPF data are of high quality providing an RV precision of 1.5\,\ms\ on AD~Leo. A total of 35 HPF RVs were obtained, five of which during HPF commissioning and 30 afterwards, namely ``HPF1'' and ``HPF2''. The HPF data in this paper are tabulated in \citetalias{Robertson2020_ADLeo_HPF}. These data overlap with optical RV data from HARPS-N and show also an amplitude decrease between observing seasons.

\section{Wavelength dependence of RV signal in CARMENES data} \label{sec:wavelengthdependence}

Signals in RV measurements induced by dark spots corotating on the stellar surface are expected to be more pronounced in the bluer wavelength regime for M-dwarf stars \citep[e.g.,][]{Desort2007_stellaractivity,Reiners2010,Mahmud2011,Sarkis2018}. In contrast, a true planetary signal should have an amplitude and phase that are consistent both in time and as a function of wavelength.

\subsection{Chromatic index} \label{sec:crx}

The chromatic index is a photospheric activity indicator that measures the RV-log $\lambda$ correlation where a straight line is fit to the RV values computed from individual \'echelle orders as a function of log $\lambda$ \citep{Zechmeister2018_SERVAL, Tal-Or_crx:2018}. This acts as a measure of wavelength dependence as caused by cool spots on the surface of M dwarfs. However, without further modeling, it does not provide any insight as to how large the spot coverage fraction is or what the star-spot temperature contrast may be. 
The RV-CRX correlations are shown for CARMENES VIS, CARMENES NIR, and HARPS, in Figs.~\ref{fig:carmvisactivity},~\ref{fig:carmvniractivity}, and~\ref{fig:harpsactivity}, respectively. The CARMENES VIS data show a clear anticorrelation ($r$ = --0.82) indicating chromatic dependency and HARPS surprisingly demonstrates a strong, positive correlation ($r$ = +0.80), whereas the CARMENES NIR data present a moderate, negative correlation ($r$ = --0.30). 

Different slopes at different wavelengths are not necessarily unexpected. In principle, this depends on which mechanism is dominating, namely either the star-spot temperature contrast or Zeeman broadening, where a more negative RV-CRX correlation suggests the former to be predominant. This is also equivalent to the amplitude of the RVs due to stellar features decreasing with longer wavelengths (Sect.~\ref{sec:wavelengthsegments}). 
The fact that the slopes of CARMENES and HARPS are in contradiction could simply indicate that different mechanisms are prevalent. Likewise, taking into consideration that we are probing very different spectral lines at different wavelengths may introduce a trend with wavelength that is not yet well understood. 
When considering only the orders of CARMENES VIS that overlap with the HARPS wavelength range, the correlation would still stay the same. Here, the wavelength range considered is now equivalent, nonetheless the signs of the slopes are inconsistent. No firm conclusion can be drawn, though this could be a result of different mechanisms dominating during different time periods of long-term stellar activity. 

Furthermore, the correlation plot for the CARMENES VIS demonstrates a ``closed-loop'' (circular) behavior (see Sect.~\ref{sec:auxdata} for a discussion). Removing the CRX trend on just the CARMENES data results in two-fold decrease in the rms of the corrected RVs (from 18.5\,\ms\ to 9.2\,\ms\ for first season and from 15.6\,\ms\ to 9.3\,\ms\ for the second). This subtraction was not performed on the CARMENES NIR data since the Pearson-$r$ coefficient was not large enough \citep{Jeffers2022}.


\subsection{Wavelength segments} \label{sec:wavelengthsegments}

The CARMENES VIS channel records 55 \'echelle orders, 42 of which are considered when computing the RV measurement via a weighted mean \citep{Zechmeister2018_SERVAL}. We chose to combine these orders into four wavelength segments where each segment consists of ten (or 11) \'echelle orders in order to preserve some precision, and the RVs are then recomputed for each respective wavelength coverage. Similarly, the CARMENES NIR channel has 28 RV orders over the $Y, J,$ and $H$ photometric bands. Due to telluric contamination, especially in the $J$ band \citep{Reiners2018b}, only a selected few orders are considered \citep{Bauer2020}. We then use two wavelength segments, one for the $Y$ and another for the $H$ band, consisting of 12 and seven individual orders, respectively. For this wavelength segment analysis, we consider just the season one CARMENES data because the data from season two exhibit an amplitude decrease (see Sect.~\ref{sec:modelresults}), and likewise, breaking up the orders would introduce noise given the few data points over a short time span.

\begin{figure}
\centering
\includegraphics[width=\hsize]{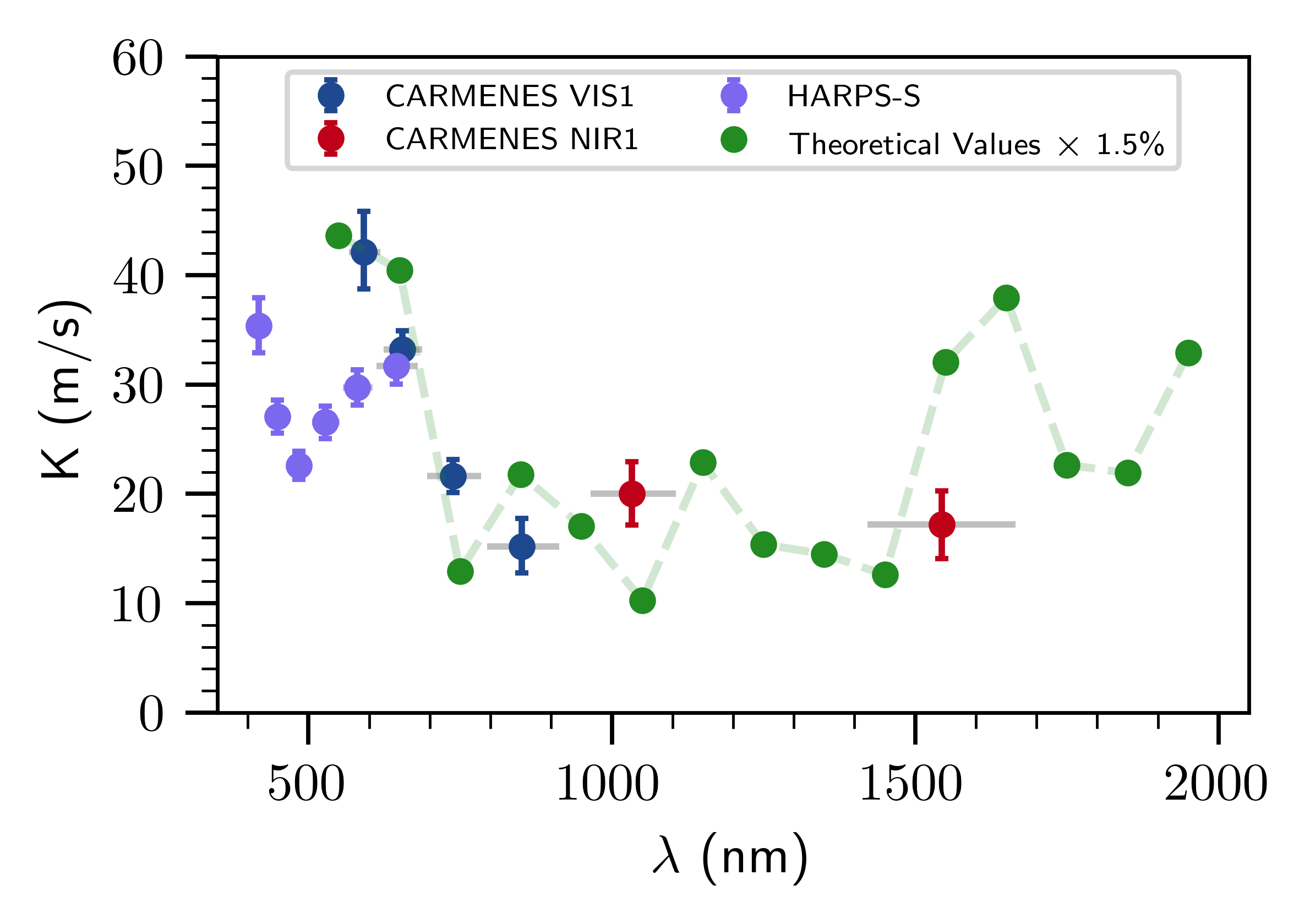}
\caption{Radial-velocity semi-amplitudes as a function of wavelength for the wavelength chunks from HARPS and the CARMENES VIS and NIR spectographs. The gray horizontal lines for each data point correspond to the wavelength coverage considered when recomputing the RV for the wavelength chunk. The green dots connected by a dashed line represent the theoretical values of a 1.5\% spot coverage on a 3700\,K, $v \sin{i}$ of 5\,k\ms\ star with a temperature difference of 200\,K taken from \cite{Reiners2010}. The theoretical values are binned in 2 to serve as a better comparison to the wavelength bins provided by the real data.}
\label{fig:Kvswavelength}
\end{figure}

Each wavelength segment is treated as an individual data set. We fit a simple sinusoid to obtain $K$, the semi-amplitude of the signal and 1$\sigma$ errors. When doing this, the semi-amplitude clearly decreases with increasing wavelength in the optical regime, but then reaches a plateau when continuing in the near-IR (Fig.~\ref{fig:Kvswavelength}). This behavior of decreasing but then constant RV semi-amplitude is in agreement with \cite{Reiners2010} for a dark spot on the surface of an M dwarf; specifically, a spot covering 1.5\% of the projected surface, with a temperature 200\,K cooler than the star (assumed $T_\textnormal{eff} = 3700$\,K), and a stellar $v \sin{i}$ of 5\,k\ms, in line with AD~Leo's stellar parameters\footnote{We adopt the higher value of $v \sin{i}$ (i.e., 5\,k\ms) rather than 2\,k\ms\ \citep[as also computed in][]{Reiners2010} because AD~Leo is relatively active for its $v \sin{i}$.} (Table~\ref{tab:stellarparams}). Simulations show a linear relation between spot coverage and RV semi-amplitude for low spot coverage values. A large spot-star temperature difference ($T_\textnormal{spot} = \frac{2}{3}T_\textnormal{eff}$), in comparison, would not cause a notable semi-amplitude dependency as a function of wavelength; however, it is not likely for cooler stars, such as AD~Leo, to have large spot-star temperature differences \citep{Bauer2018_starspots}.

Likewise, the HARPS instrument covers 72 spectral orders, of which 61 produce reliable RVs after being processed by \serval\ (signal-to-noise is too low for the others). Similarly, we computed six wavelength chunks with ten spectral orders each (11 for the reddest chunk) and followed the same methodology as for the CARMENES wavelength chunks. 
\cite{Reiners2013} and \citetalias{Tuomi2018_ADLeo} performed similar analyses but differed in interpretation. \citetalias{Tuomi2018_ADLeo} suggested there is no dependence on wavelength, whereas \cite{Reiners2013} claimed otherwise and mentioned that this is the first case of a known active star with increasing amplitude with wavelength.
Here, we find a slight positive incline, which is plotted for comparison to the CARMENES data in Fig.~\ref{fig:Kvswavelength}. The positive slope can be interpreted as being due to the Zeeman effect, which has the opposite effect compared to a spot-temperature difference where the RV amplitude is predicted to increase for redder wavelengths \citep{Reiners2013}. 
For the overlapping wavelengths, the amplitudes of the wavelength chunks do not particularly agree which can simply be an artifact that HARPS data were taken at a time where AD~Leo exhibited less activity in the RVs, as is likewise present behavior in the photometry (Sect.~\ref{sec:phot}).

\section{Identifying individual spectral lines affected by stellar activity} 
\label{sec:spectrallines}

We further investigate the possible effect of stellar activity on the spectral lines themselves. We follow an approach similar to \cite{Dumusque2018_spectrallines}, which has proven to be fruitful for a selection of G-K dwarfs whose RVs are dominated by activity. Essentially, we use the CCF technique, where we obtain a binary mask that contains all available, reliable lines for a spectrum \citep[following][]{Lafarga2020}. We then compute individual RVs of the few thousands of spectral lines we have identified, and obtain an RV time series for each line. To classify the lines according to their sensitivity to activity, we correlate their RVs to an activity indicator obtained from the same spectra (such as the CRX, BIS, or the total RV). We then select a subsample of spectral lines that are least affected by stellar activity (those that do not show a strong correlation) and recompute the RVs using this subsample to mitigate stellar activity. The recomputed RVs then have a smaller scatter and the modulation due to stellar rotation decreases.

This technique has been tested for other M dwarfs similar to AD~Leo (spectral types 3.0 to 4.5\,V, relatively low rotational velocities and high activity levels) that are well-known to exhibit strong stellar activity signals (e.g., YZ~CMi, EV~Lac) and appears to perform well and as expected (Lafarga priv. comm.).
We found similar results regardless of the activity indicator (total RV, CRX or BIS) used to compute the correlations with the individual line RVs, with the total RV yielding slightly smaller RV scatters.

\subsection{Individual spectral lines  using the CARMENES data}
\begin{figure*}
    \centering
    \includegraphics[width=2\columnwidth]{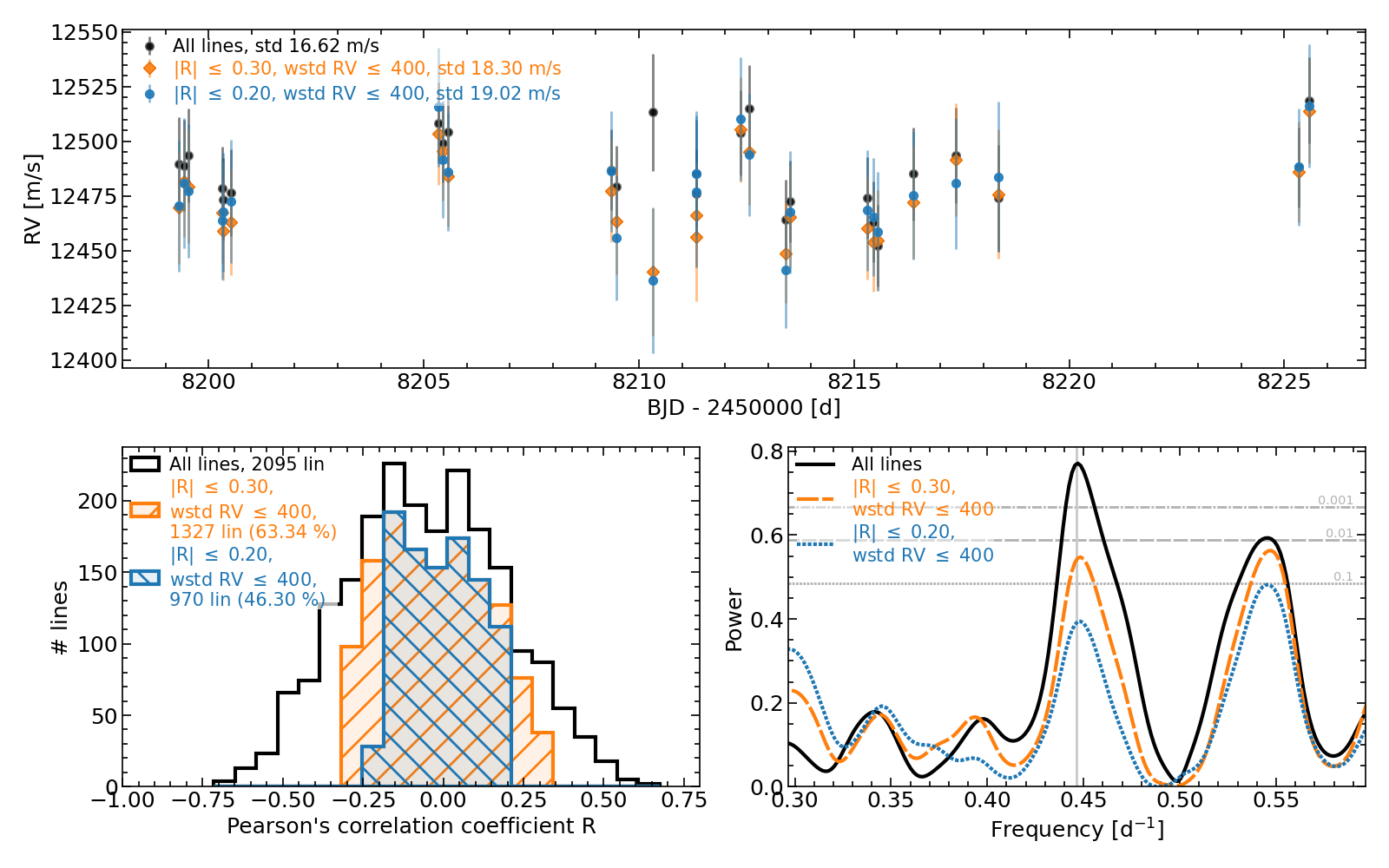}
    \caption{Three various subsamples of the spectral lines: black -- all, orange -- $\abs{r} \leq 0.30$, and blue -- $\abs{r} \leq 0.20$. \textit{Top:} Absolute RVs versus time for the first CARMENES VIS season. \textit{Bottom left:} Histogram of the Pearson-$r$ correlation coefficient when comparing the RVs from individual lines to the CRX. The criteria for the subsamples are illustrated here. \textit{Bottom right:} Zoomed-in GLS periodograms around the signal of interest of the RVs of the subsamples. The orange subsample still shows power at 2.23\,d, whereas the blue subsample still has a nonsignificant bump but it is noise limited.}
    \label{fig:spectrallines}
\end{figure*}

Specifically for AD~Leo, we compare the computed individual line RVs to the CRX or the BIS and estimate the strength of stellar activity based on these correlations.
We focus on just the first season in 2018, namely, VIS1, since the second season only covers a bit more than one rotation cycle and this then introduces too much scatter due to the photon noise being too high.
To quantify the correlation strength, we used the Pearson correlation coefficient $r$. We considered lines as `inactive' with $r$ close to 0. We also discarded lines that had time series scatter larger than 400\,\ms, as measured from their weighted standard deviation (WSTD) RV; these are weak lines that mostly add noise to the recomputed RVs.
Our results for the correlation with the CRX are shown in Fig.~\ref{fig:spectrallines}, where we generated three line subsamples: (1) all spectral lines -- black, (2) where $\abs{r} \leq 0.30$ -- orange
, and (3) where $\abs{r} \leq 0.20$ -- blue. We obtained similar values using the correlation with the BIS. We recomputed the RVs from both the orange and blue subsamples, thus using the least activity-affected lines. Assuming the scatter of the RV data to be dominated by the stellar activity, we then would expect a decrease in the time series RV scatter, which we do not observe.
Regarding the modulation, the periodograms of these two data sets show a decrease in the power of the 2.23\,d peak compared to the all-lines data set, but there is still some power left (false alarm probability (FAP) > 10\,\% and almost 10\,\% for the orange and blue subsets respectively).

This would be clear if the RV scatter decreased for the orange or blue data sets. However, since we do not observe such a decrease, it is difficult to discern what is causing the periodogram behavior. The increase in the RV scatter could be caused by increasing photon noise in the RVs of the orange and blue data sets (because we are using a smaller number of lines to compute the RVs, hence less signal). Then, the decrease in the periodogram power could be due to this increasing noise, and not to a decrease in the activity signal.

Compared to the results from other stars (e.g., EV~Lac, YZ~CMi), the correlations found for AD~Leo are much weaker (the mean correlation coefficient $r$ is $\sim$ --0.2 and some of them show $\abs{r}$ $\geq$ 0.8 for the correlation with the CRX, while for AD~Leo, the mean is at 0 and very few lines have $\abs{r}$ $\geq$ 0.6). This lack of clear correlations indicates that the correlations that we find for AD~Leo do not have much information related to the activity of the star, and this could be why we are not able to effectively mitigate the stellar activity signal in the RVs.

This difference in the correlation strength could be due to the different RV amplitudes of the stars. AD~Leo shows a small RV amplitude compared to the other considered stars: $K$ $\sim$ 25\,\ms\ in comparison to EV~Lac with $\sim$100\,\ms.
Both stars have similar spectral types and activity levels \citep[for EV~Lac, pEW(H$\alpha$) = --4.983\,$\pm$\,0.021, as computed from the CARMENES observations,][]{Schoefer2019}, so the difference in RV amplitudes seems to be caused by different spot configurations.
AD~Leo has a relatively low inclination ($i \sim 13^{\circ}$) in comparison to EV~Lac or the other considered stars \citep[$\geq 60^{\circ}$, see e.g.,][]{Morin2008}. This close to pole-on inclination could cause any visible corotating spots to induce a smaller modulation in the RVs simply because $v \sin{i}$ is smaller. Also, the photosphere of AD~Leo could be more homogeneously spotted, also inducing smaller RV modulations.

To summarize, we recomputed the RVs using only the lines least affected by activity in the AD~Leo spectra, and observed a decrease in the periodogram peak at 2.23\,d (0.1\,\% to 10\,\%); however, there was still some significant residual power and we did not observe a significant decrease in the RV scatter. This could be due to AD~Leo having a different activity signal in the RVs than other stars with similar characteristics, for which we observe a clear mitigation of activity. 

\subsection{Other spectroscopic activity indicators} 
\label{sec:otheractivityindicators}
To aid in disentangling the origin of a signal, a variety of stellar activity indicators, such as photometric variability, bisector inverse slope (BIS, the measurement of asymmetry of the CCF), Ca~\textsc{ii}~infrared triplet (IRT), H$\alpha$, CRX, the dLW, or the TiO bands, has been shown to be successful in identifying and indicating a nonplanetary signal.  
In the literature, periodic RV signals are often considered planetary in nature if the same signal is not found in activity indicators. Therefore, it is important to examine as many activity indicators as possible.

Correlations between the mentioned stellar activity indicators (when available) and the RVs along with the GLS periodograms for these indicators can be found in Appendix~\ref{sec:activityappendix} for both the CARMENES VIS and NIR channel as well as for the HARPS instrument. The strong correlations found with the CRX was already discussed in Sect.~\ref{sec:crx}. 
The BIS-RV anticorrelation for the VIS channel ($r$ = --0.79) is inline with the same anticorrelation found by \citetalias{Carleo2020_ADLeo_GIARPS} in the HARPS-N data ($r$ = --0.74).
Additionally, we look at the dLW 
However, in our case, there seems to be no correlation. Along with the dLW, other expected stellar activity indicators such as the emission of H$\alpha$ \citep{Kuerster2003, Hatzes2015, Hatzes2016, Jeffers2018, Barnes2014} and the Ca~\textsc{ii}~IRT \citep{GomesdaSilva2011, Martin2017, Robertson2015, Robertson2016} seem to not show peaks at 2.23\,d in periodograms and exhibit no strong or moderate correlations.
These findings are in agreement with \cite{Lafarga2021} for a high-mass and H$\alpha$-active star as AD~Leo. In addition, we also consider the pseudo-equivalent widths (pEW') of various potential stellar activity lines
as described in \cite{Schoefer2019}, though find no correlations or significant peaks other than suggestive bumps at 1.8\,d (i.e., the daily alias of 2.23\,d).
 
\section{Modeling the time dependence of the RV signal}
\label{sec:timedependence}
Discontinuous sampling over the course of a long time baseline impedes obtaining key information as to what occurs in large data gaps --- as is the case between the majority of the HIRES, HARPS data and CARMENES, GIARPS, HPF data for AD~Leo being 12 years apart. Remarkably, the 2.23\,d signal persisted over all this time, however, it is evident that amplitude changes and phaseshifts occurred. Therefore, given such a rich RV data set, we test out the current approaches for modeling stellar activity on a real-life example and present our results below.

\subsection{Model setup} 

We model the RV data using two components: a stable component (in amplitude, period, and phase) and a variable component. The former is described by a deterministic model, typically a circular or eccentric Keplerian orbit. We would like to emphasize that when we use a Keplerian model, we consider it more as a ``stable model'' which may include both a planetary companion and a stable aspect of stellar activity, or only one of the two. Therefore, parameters that are typically called $P_\textnormal{planet}$ and $K_\textnormal{planet}$ to represent the period and semi-amplitude of the Keplerian signal are rather referred as $P_\textnormal{stable}$ and $K_\textnormal{stable}$ to indicate that we are considering a ``stable mode'' that appears persistent. From here on, we use the terms Keplerian model and stable model interchangeably, though we are not claiming that this component is solely due to a planetary companion. As for the variable component, typically comprising the quasi-periodic behavior of stellar activity, we do not currently possess a deterministic model, therefore, we use a nonparametric GP model to describe these modulations. 
We employ two different kernels.

The first is an exponential-squared-sine-squared kernel, that is, quasi-periodic (QP-GP) kernel, provided by \george\ \citep{george}, 

\begin{equation} \label{eqn:quasiperiodickernel}
k_{i,j}(\tau)=\sigma^2_\textnormal{GP}\exp\left[-\alpha_\textnormal{GP}\tau^2-\Gamma_\textnormal{GP}\sin^2\left(\frac{\pi\tau}{P_\textnormal{GP,\ rot}}\right)\right]
\end{equation}

\noindent where $\tau = |t_{i} - t_{j}|$ is temporal distance between two points, $\sigma_\textnormal{GP}$ is the amplitude of the GP modulation, $\alpha_\textnormal{GP}$ is the inverse length-scale\footnote{The parameter $\alpha_\textnormal{GP}$ is defined as $1/l^2$, where $l$ is the timescale of variations. The original relation in \citet{juliet} incorrectly defined $\alpha = 1/2l^2$. This has since been corrected.} of the GP exponential component, $P_\textnormal{GP,\,rot}$ corresponds to the recurrence timescale, and $\Gamma_\textnormal{GP}$ is the smoothing parameter. The former term is an exponential that can model the decorrelation due to the changes in phase and amplitude as active regions grow and decay over time, whereas the latter term accounts for the reoccuring periodicity.

The second kernel is a sum of two stochastically driven, damped harmonic oscillator (SHO) terms, or a double SHO GP (dSHO-GP), where the power spectrum of one SHO term is given by \cite{Anderson1990}, 

\begin{subequations} \label{eq:dsho}
\begin{align}
    \textnormal{SHO}_1(\omega_\textnormal{GP}) = \sqrt{\frac{2}{\pi}} \frac{S_0\omega_1^4}{(\omega^2_\textnormal{GP}-\omega^2_1)^2 + \omega^2_1\omega_\textnormal{GP}^2/Q^2_1}
\end{align}\\
\textnormal{and}\\
\begin{align}
    \textnormal{SHO}_2(\omega_\textnormal{GP}) = \sqrt{\frac{2}{\pi}} \frac{S_0\omega_2^4}{(\omega^2_\textnormal{GP}-\omega^2_2)^2 + \omega^2_2\omega_\textnormal{GP}^2/Q^2_2},
\end{align}\\
\textnormal{for which we applied a reparametrization using the hyperparameters}\\
\begin{align}
    Q_1      & = 0.5 + Q_0 + \delta Q\\
    \omega_1 & = \frac{4\pi Q_1}{P_\textnormal{GP,\,rot}\sqrt{4Q_1^2 -1}}\\
    S_1     & = \frac{\sigma_\textnormal{GP}^2}{(1+f)\omega_1 Q_1}\\
    Q_2     & = 0.5 + Q_0\\
    \omega_2 & = 2\omega_1  = \frac{8\pi Q_1}{P_\textnormal{GP,\,rot}\sqrt{4Q_1^2 -1}}\\
    S_2 & = \frac{f \sigma_\textnormal{GP}^2}{(1+f)\omega_2 Q_2},
\end{align}
\end{subequations}

\noindent and where $\sigma_\textnormal{GP}$ is again the amplitude of the GP kernel, $P_\textnormal{GP,\,rot}$ is the primary period of the variability, $Q_0$ is the quality factor for the secondary oscillation, $\delta Q$ is the difference between the quality factors of the first and second oscillations, and $f$ represents the fractional amplitude of the secondary oscillation with respect to the primary one.

We investigate multiple models to see which one is preferred for each individual data set as well as for the combined data set comprising all available RVs. The models being tested are: 

\begin{enumerate}
    \item Keplerian-only models, i.e, both circular and eccentric
    \item GP-only model, either with a QP-GP or a dSHO-GP, as a proxy to describe the quasi-periodic nature of the stellar activity
    \item Mixed (Keplerian + GP) models, to describe stable components with variable ones 
\end{enumerate}

We use the model-fitting python package, \juliet\ \citep{juliet} in order to compare these models using a Bayesian framework. For our purposes, the RVs are modeled by \texttt{radvel}\footnote{\url{https://radvel.readthedocs.io/en/latest/}} \citep{radvel}, and the GP models with the help of \texttt{george}\footnote{\url{https://george.readthedocs.io/en/latest/}} \citep{george} and \texttt{celerite}\footnote{\url{https://celerite.readthedocs.io/en/stable/}} \citep{celerite}, where we use the \texttt{dynesty}\footnote{\url{https://github.com/joshspeagle/dynesty}} package \citep{dynesty,dynesty2020} to execute the Nested Sampling algorithm in order to efficiently compute the Bayesian model log evidence, $\ln \mathcal{Z}$. 
The main motivation for calculating the Bayesian log evidence ($\ln \mathcal{Z}$) is to perform model comparisons. Outlined by \cite{Trotta2008}, we follow the rule of thumb that a $\Delta \ln \mathcal{Z}$ greater than 5 between two models indicates strong evidence in favor for the model with the larger Bayesian log evidence (odds are $\sim$150 to 1), whereas a $\Delta \ln \mathcal{Z}$ more than 2.5 indicates moderate evidence for the winning model, and anything less we consider to be inconclusive.

\subsection{Prior setup} \label{sec:priorsetup}
The RV periodograms (Fig.~\ref{fig:periodogramrv}) indicate the interesting signal to be around 2.23\,d. Thus, for all models with a Keplerian, the prior of the period of the stable component, $P_\textnormal{stable}$, was kept uniform, but relatively narrow, $\mathcal{U}$(2.0\,d, 2.5\,d), in order to avoid picking up aliases (i.e., 1.8\,d due to the daily sampling), and similarly for the time of transit center\footnote{The time-of-transit center is used within the \juliet\ framework in regards to the phase of the orbit.}, $t_0$, $\mathcal{U}$(2458200\,d, 2458202\,d), in order to cover solely one cycle and to avoid picking up other potential modes. The semi-amplitude prior was simply uniform as well, $\mathcal{U}$(0\,\ms, 50\,\ms).

As for the GP models, the rotational period, $P_\textnormal{GP,\ rot}$ follows the same prior as $P_\textnormal{stable}$ for consistency. Then, we consider wide log-uniform priors for the other QP-GP hyperparameters: 
$\Gamma_\textnormal{GP}$ (between $10^{-2}$ and $10^{2}$) and $\alpha_\textnormal{GP}$ (between $10^{-10}$\,\,d$^{-2}$ and $10^{-1}$\,\,d$^{-2}$, corresponding to timescales of $\sim$3\,d -- $\sim$70\,000\,d).
For the dSHO-GP hyperparameters: $Q_0$ was log-uniform (from $10^2$ to $10^5$), as well as $\delta Q$ (from $10^{-1}$ to $10^5$), and $f$ was kept uniform (between $0$ and $1$).
For both GP kernels, the sigma of the GP was kept log-uniform (between 0.1 and $70$\,\ms) where each instrument had its own, noted as $\sigma_\textnormal{GP,\ inst}$, because each instrument has its own characteristics, such as noise level, zero point offset, wavelength range, or intrinsic stellar jitter. The other GP hyperparameters are shared. Additional instrumental jitter terms (log-uniform from $10^{-2}$ to $30$\,\ms) and offsets for each individual instrument were considered as well. 

\subsection{Results} \label{sec:modelresults}
We applied this recipe to the three following data sets: CARMENES VIS only, CARMENES NIR only, and the entire RV data set. The resulting posteriors on the $P_\textnormal{stable}$ ($P_\textnormal{GP,\ rot}$ if using a GP), semi-amplitude $K_\textnormal{stable}$ ($\sigma_\textnormal{GP,\ inst}$ if GP), added instrumental jitter terms ($\sigma_\textnormal{inst}$) and eccentricity along with the differences in Bayesian log evidence can be found in Table~\ref{tab:models}.

\subsubsection{CARMENES only}

Focusing on just the CARMENES season one data, an eccentric ($e\sim0.19$) Keplerian model is preferred over a circular model for the VIS channel ($\Delta \ln \mathcal{Z} \sim 5$). This is in agreement with the optical HARPS-N data where \citetalias{Carleo2020_ADLeo_GIARPS} also found a similar eccentricity when applying a Keplerian-only model. 
As for the NIR channel, a circular model performs best. Moreover, the NIR data does not actually have a clear model preference, simply attributed to the lower precision. 
The amplitude decreased from $28.43^{+0.94}_{-0.98}$\,\ms\ in the VIS1 data to $18.0^{+2.6}_{-2.7}$\,\ms\ in the NIR1 data.
When introducing combined stable plus GP models, the stable component dominates and the GP component is not needed, in other words, $\sigma_\textnormal{GP}$ becomes consistent with zero. This finding is anticipated given that the time span of the data is roughly $\sim$10 cycles of the periodicity, which would be too short for any noticeable changes of the stellar spot pattern assuming a long spot lifetime.

For the season two data, we slightly readjusted the priors as introduced in Sect.~\ref{sec:priorsetup}. The time of transit center was moved to $\mathcal{U}$(2458893\,d, 2458895\,d) to comply with the pertinent time stamps. Likewise, the period was narrowly constrained, $\mathcal{U}$(2.22\,d, 2.24\,d), to ensure the correct periodicity is recognized given the short time baseline of less than two periodic cycles. Likewise for this reason, just the circular Keplerian-only models were performed, and as a result, the posterior values are not included in Table~\ref{tab:models}.
Similarly to the season one data, there was also an amplitude decrease between the instruments, namely from $17.02^{+0.63}_{-0.62}$\,\ms\ to $9.2^{+3.0}_{-2.6}$\,\ms\ between VIS2 and NIR2, respectively, both of which are lower with respect to their season one counterparts. Figure~\ref{fig:carmrv} shows how the time series and phase-folded plots using the best model fits for the four CARMENES data sets appear reasonable, showing a uniformly distributed scatter in the model residuals. We additionally tested out combining the two seasons with each respective instrument (i.e., VIS1+VIS2, NIR1+NIR2) and applied all the models (including those with GPs), finding that the circular Keplerian was favored for both.


\subsubsection{Whole data set}
Next we considered the whole AD~Leo RV data set. 
We expect the Keplerian-only models, where the amplitude for all given instruments is shared, to perform poorly given that the amplitude is clearly decreasing as a function of wavelength (shown in Sect.~\ref{sec:wavelengthdependence}).
The idea of combining all available data sets covering a wide wavelength range is that the data set with the smallest amplitude, specifically the NIR data, will act as an upper limit for any stable signal, whereas the GP component will adapt the difference. Within the framework of \juliet, the data is first attempted to fit the deterministic model as best as possible, where the GP is then applied to the residuals. Likewise, we expect that the added jitter terms and GP amplitude parameters for the redder instruments will be consistent with zero in the mixed model because the stable model will most likely saturate for the redder instruments. The role of the GP is to account for the excess amplitude observed with the bluer instruments.

The preferred model among all is the \textbf{mixed circular Keplerian + dSHO-GP model}. When compared to a Keplerian-only model, it exceeds in Bayesian log evidence tremendously ($\Delta \ln \mathcal{Z} \sim 180$), but it does not prevail against a dSHO-GP-only model that enormously ($\Delta \ln \mathcal{Z} \sim 3.5$). As anticipated, the Keplerian-only models fail to explain the data in the sense that finding a common amplitude is not feasible such that the jitter values become too high, and likewise, the phase shifts are quite strong that a simple sinusoid can not describe this behavior. Surprisingly, even though the dSHO-GP-only model is flexible and could model the data well, the evidence suggests that a model containing an additional stable periodic component is favored. Likewise, the same concept applies for the models including the QP-GP kernel, though including the additional stable component produced a $\Delta \ln \mathcal{Z} \sim 10$ rather than $\sim 3.5$ for the dSHO-GP models, that is, very strong evidence for one model versus bordering just moderate evidence. This delicate boundary could change the interpretation. Tests on simulated RV data based on StarSim \citep[][]{Herrero2016_starsim} indicate that a Keplerian signal is in rare cases more efficient than the QP-GP in modeling a coherent activity signal \citep{Stock2022}. 

\begin{figure*}[hbt!]
\centering
\includegraphics[width=0.45\hsize]{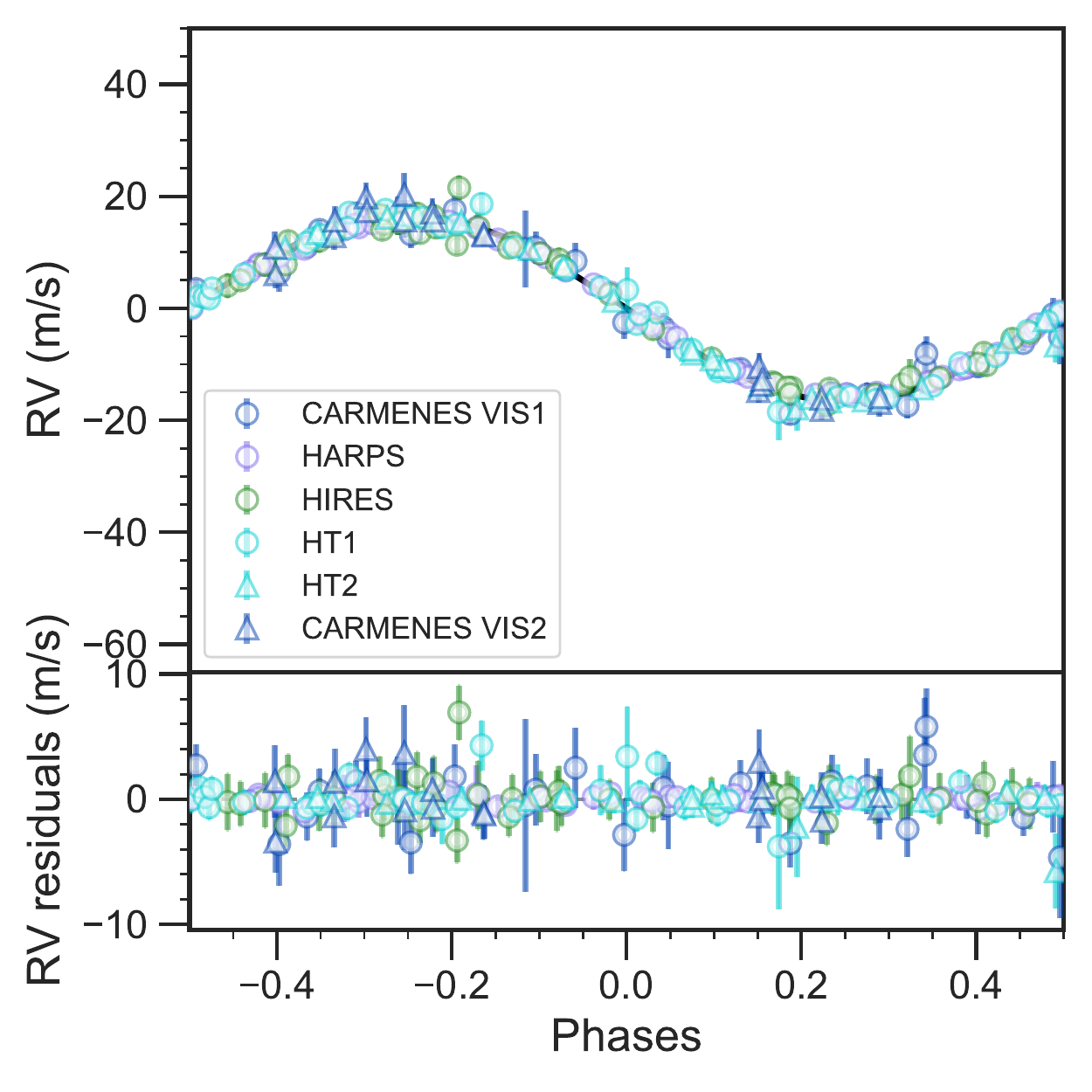}
\includegraphics[width=0.45\hsize]{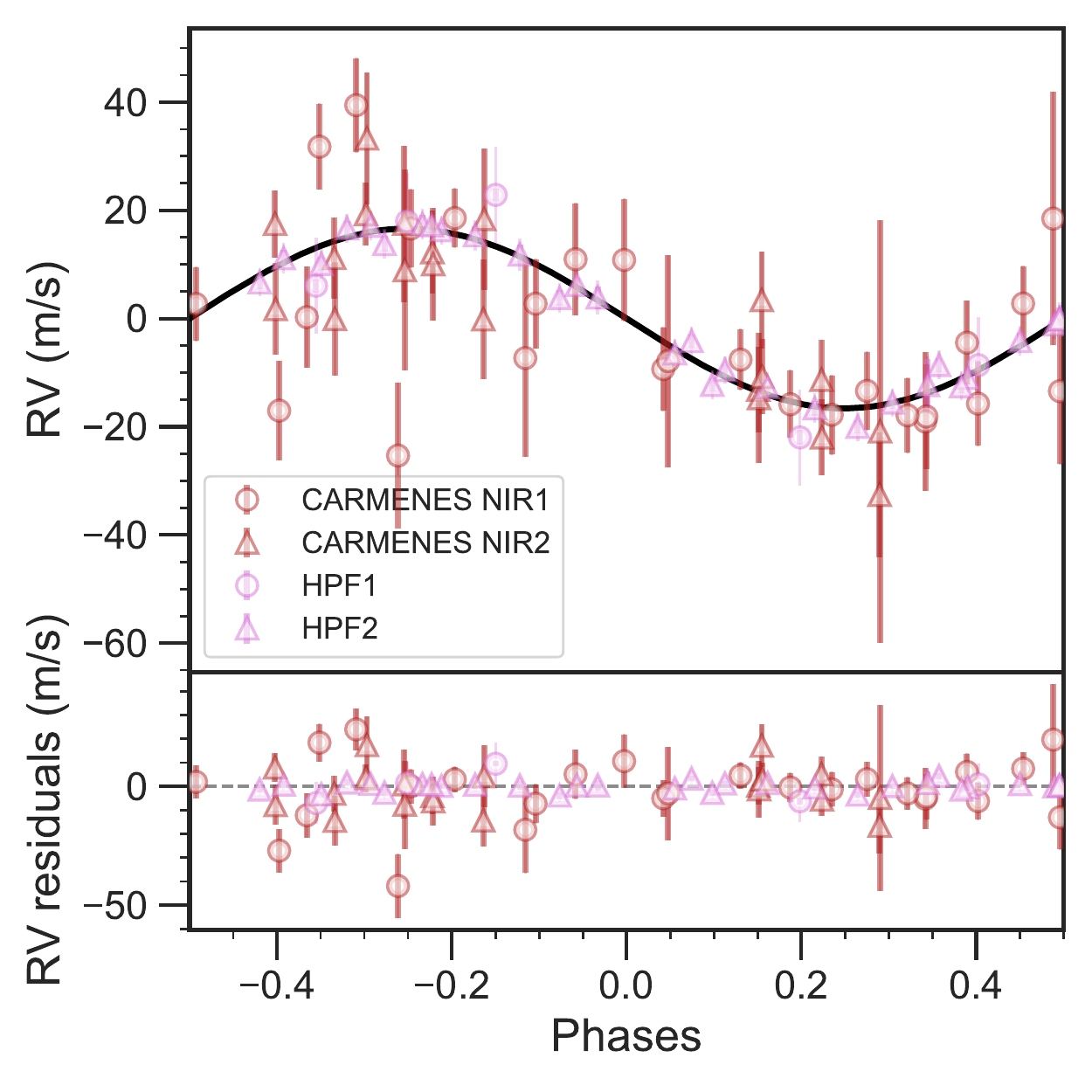}
\caption{Phase-folded plots for the optical (\textit{left}) and near-infrared (\textit{right}) instruments using the circular Keplerian + dSHO-GP model after subtracting the GP component out. The first subset of any given instrument is represented with a circle, whereas the second subset by a triangle, when applicable. The y-axis ranges are consistent between the two plots, except for the residual plots to better visualize the scatter around the fit.}
\label{fig:phasefoldrv}
\end{figure*}

We find that the stable signal amplitude $K_\textnormal{stable}$ is $16.6 \pm 2.2$\,\ms, where both the $P_\textnormal{stable}$ and $P_\textnormal{GP}$ is rounded to 2.23\,d. All posteriors for the different models can be found in Table~\ref{tab:models} and the phase-folded plots for the best model, namely the mixed model, are shown in Fig.~\ref{fig:phasefoldrv}.

To summarize our findings, we have the following takeaway points:
\begin{itemize}
    \item The CARMENES-VIS1 data alone prefer an eccentric Keplerian ($e\sim0.19$) model. This is in agreement with the HARPS-N data from \citetalias{Carleo2020_ADLeo_GIARPS}. Whereas, the CARMENES-NIR1 data alone prefer a circular Keplerian, that is, a sinusoid.
    \item The amplitude between the two CARMENES seasons decreased in both instruments. Namely, from $28.43^{+0.94}_{-0.98}$\,\ms\ to $17.02^{+0.63}_{-0.62}$\,\ms\ in the VIS channel, and from $18.0^{+2.6}_{-2.7}$\,\ms\ to $9.2^{+3.0}_{-2.6}$\,\ms\ in the NIR channel.
    \item Combining both seasons of the CARMENES data (i.e., VIS1+VIS2 and NIR1+NIR2), a circular Keplerian is preferred.
    \item Combining all data sets, the mixed circular Keplerian + GP model provides the best fit, for either GP used (QP-GP or dSHO-GP). This signifies that there is a stable and variable component in the data. 
    \item The $K_\textnormal{stable}$ of the preferred model (mixed circular Keplerian + dSHO-GP model) is $16.6 \pm 2.2$\,\ms. The amplitude of the GP component ($\sigma_\textnormal{GP,\ inst}$) is consistent with 0 for the near-IR instruments, except for HPF2 ($\sigma_\textnormal{GP,\ HPF2}= 7.8\pm 2.6$\,\ms). For the optical instruments, the earlier data, namely with HARPS, HIRES, have similar sigmas ($\sigma_\textnormal{GP,\ inst} \sim 20$\,\ms) as well as the overlapping data, with VIS1 and HT1 ($\sigma_\textnormal{GP,\ inst} \sim 13$\,\ms). For VIS2, it was consistent with zero and for HT2, it was higher with $\sigma_\textnormal{GP,\ inst} \sim 22$\,\ms.
   \item The difference in evidence between a mixed model and a GP-only model depended on the GP kernel choice (i.e., $\Delta \ln \mathcal{Z} \sim 3.5$ and $\Delta \ln \mathcal{Z} \sim 10$ when considering a dSHO-GP and QP-GP kernel, repspectively.)
\end{itemize}

\begin{table*}[hbt!]
    \centering
    \scriptsize
    \caption{Model comparison of RV fits done with \juliet\ comparing a Keplerian model, a red-noise model, and a mixture of both for the various instrument data sets: CARMENES VIS1, CARMENES NIR1, and the combined data set. The selected best model is boldfaced (refer to Sect.~\ref{sec:timedependence} for details on the priors used, model choice, and discussion of results). The columns refer to which model was being used, whereas the rows correspond to the model parameter. A larger, positive $\Delta\ln \mathcal{Z}$ value indicates a better model.}  
    \label{tab:models}
    \begin{tabular}{lcccccccc}

\hline
\hline
\noalign{\smallskip}
&   &  & \multicolumn{3}{c}{QP-GP} & \multicolumn{3}{c}{dSHO-GP} \\
\cmidrule(lr){4-6}
\cmidrule(lr){7-9}
&  circ. &  ecc. & GP &  circ. + GP &  ecc. + GP & GP &  circ. + GP &  ecc. + GP \\
\noalign{\smallskip}
\hline
\noalign{\smallskip}
CARMENES VIS1 \\

~~~$P_\textnormal{stable}$ & $2.2362^{+0.0054}_{-0.0056}$ & $2.2344^{+0.0029}_{-0.0030}$ & \ldots & $2.238^{+0.150}_{-0.014}$ & $2.2346^{+0.0033}_{-0.0035}$ & \ldots & $2.237^{+0.015}_{-0.008}$ & $2.2348^{+0.0036}_{-0.0036}$ \\
~~~$K_\textnormal{stable}$ & $25.9^{+1.4}_{-1.5}$ & $28.43^{+0.94}_{-0.98}$ & \ldots & $19^{+11}_{-16}$ & $28.4^{+1.1}_{-1.2}$ & \ldots & $22.6^{+7.2}_{-16.0}$ & $28.3^{+1.2}_{-1.5}$  \\
~~~$P_\textnormal{GP}$ & \ldots & \ldots & $2.2332^{+0.0035}_{-0.0038}$ & $2.226^{+0.011}_{-0.019}$ & $2.28^{+0.16}_{-0.19}$ & $2.2329^{+0.0048}_{-0.0049}$ & $2.224^{+0.013}_{-0.017}$ & $2.24^{+0.19}_{-0.14}$\\
~~~ecc. & 0.0 (fixed) & $0.192^{+0.030}_{-0.031}$ & \ldots & 0.0 (fixed) & $0.186^{+0.032}_{-0.033}$ & \ldots & 0.0 (fixed) & $0.187^{+0.034}_{-0.038}$ \\
~~~$\sigma_\textnormal{GP, CARMENES\ VIS1}$ & \ldots & \ldots & $38^{+18}_{-14}$ & $23^{+24}_{-15}$ & $1.8^{+7.6}_{-1.7}$ & $18.3^{+8.2}_{-5.1}$ & $9.7^{+7.7}_{-4.2}$ & $0.79^{+3.00}_{-0.74}$ \\
~~~$\sigma_\textnormal{CARMENES\ VIS1}$ & $4.31^{+1.00}_{-0.83}$ & $1.2^{+1.1}_{-1.1}$ & $1.2^{+1.1}_{-1.1}$ & $0.45^{+1.40}_{-0.42}$ & $0.6^{+1.4}_{-0.6}$ & $0.26^{+1.10}_{-0.23}$ & $0.27^{+1.20}_{-0.24}$ & $0.61^{+1.40}_{-0.57}$  \\
\noalign{\smallskip}
\hline
\noalign{\smallskip}
~~~$\ln \mathcal{Z}$ & --94.06 & \textbf{--87.71} & --90.06 & --90.39 & --87.08 & --87.86 & --88.18 & --87.05\\
~$\Delta\ln \mathcal{Z}$ & --6.35 & \textbf{0.0} & --2.35 & --2.68 & 0.64 & --0.14 & --0.46 & 0.66 \\ 
\noalign{\smallskip}
\hline
\noalign{\medskip}
\hline
\noalign{\smallskip}
CARMENES NIR1 \\

~~~$P_\textnormal{stable}$ & $2.269^{+0.016}_{-0.017}$ & $2.286^{+0.016}_{-0.014}$ & \ldots & $2.264^{+0.049}_{-0.110}$ & $2.287^{+0.023}_{-0.019}$ & \ldots & $2.270^{+0.051}_{-0.110}$ & $2.288^{+0.025}_{-0.019}$  \\
~~~$K_\textnormal{stable}$ & $18.0^{+2.6}_{-2.7}$ & $21.8^{+3.9}_{-3.5}$ & \ldots & $16.3^{+4.1}_{-5.2}$ & $20.7^{+4.3}_{-5.0}$ & \ldots & $16.6^{+4.0}_{-4.7}$ & $20.9^{+4.2}_{-4.7}$ \\
~~~ecc. & 0.0 (fixed) & $0.488^{+0.078}_{-0.130}$ & \ldots & 0.0 (fixed) & $0.470^{+0.092}_{-0.200}$ & \ldots & 0.0 (fixed) & $0.471^{+0.091}_{-0.190}$  \\
~~~$P_\textnormal{GP}$ & \ldots & \ldots & $2.276^{+0.023}_{-0.042}$ & $2.19^{+0.19}_{-0.11}$ & $2.22^{+0.20}_{-0.15}$ & $2.286^{+0.031}_{-0.031}$ & $2.174^{+0.200}_{-0.097}$ & $2.22^{+0.21}_{-0.15}$  \\
~~~$\sigma_\textnormal{GP, CARMENES\ NIR1}$ & \ldots & \ldots $20.9^{+16.0}_{-7.4}$ & $16^{+21}_{-11}$ & $4.4^{+17.0}_{-4.3}$ & $17.1^{+12.0}_{-6.8}$ & $7.5^{+7.8}_{-6.2}$ & $2.6^{+6.9}_{-2.5}$ \\
~~~$\sigma_\textnormal{CARMENES\ NIR1}$ & $0.97^{+5.30}_{-0.93}$ & $0.34^{+2.80}_{-0.31}$ & $0.28^{+2.30}_{-0.25}$ & $0.30^{+2.70}_{-0.26}$ & $0.25^{+2.10}_{-0.23}$ & $0.43^{+4.10}_{-0.39}$ & $0.31^{+2.90}_{-0.28}$ & $0.26^{+2.20}_{-0.23}$\\
\noalign{\smallskip}
\hline
\noalign{\smallskip}
~~~$\ln \mathcal{Z}$ & \textbf{--109.92} & --108.95 & --109.88 & --108.72 & --108.30 & --111.07 & --108.98 & --108.32\\
~$\Delta\ln \mathcal{Z}$ & \textbf{0.0} & 0.97 & 0.04 & 1.19 & 1.62 & --1.15 & 0.94 & 1.60\\
\noalign{\smallskip}
\hline
\noalign{\medskip}
\hline
\noalign{\smallskip}
\multicolumn{6}{l}{HIRES + HARPS + CARMENES VIS + CARMENES NIR + HARPS-N + HPF} \\
\noalign{\smallskip}
~~~$P_\textnormal{stable}$ & $2.226228^{+0.000028}_{-0.000026}$ & $2.227311^{+0.000018}_{-0.000019}$ & \ldots & $2.226329^{+0.000078}_{-0.000066}$ & $2.226364^{+0.000900}_{-0.000088}$ & \ldots & $2.226345^{+0.000098}_{-0.000084}$ & $2.22638^{+0.00095}_{-0.00010}$ \\
~~~$K_\textnormal{stable}$ & $17.13^{+0.59}_{-0.58}$ & $28.3^{+0.9}_{-1.0}$ & \ldots & $15.7^{+1.3}_{-4.5}$ & $12.9^{+3.7}_{-2.7}$ & \ldots & $16.57^{+0.85}_{-2.20}$ & $14.6^{+2.5}_{-4.0}$ \\
~~~$P_\textnormal{GP}$ & \ldots & \ldots & $2.2245^{+0.0021}_{-0.0021}$ & $2.2250^{+0.0022}_{-0.0021}$ & $2.2252^{+0.0023}_{-0.0021}$ & $2.22775^{+0.00054}_{-0.00067}$ & $2.2270^{+0.0010}_{-0.0011}$ & $2.22735^{+0.00095}_{-0.00120}$ \\
~~~ecc. & \ldots & $0.183^{+0.022}_{-0.023}$ & \ldots & \ldots & $0.072^{+0.100}_{-0.051}$ & \ldots & \ldots & $0.072^{+0.099}_{-0.052}$ \\
~~~$\sigma_\textnormal{GP, HIRES}$ & \ldots & \ldots & $25.6^{+4.6}_{-3.5}$ & $20.8^{+3.6}_{-2.8}$ & $21.1^{+3.7}_{-2.9}$ & $23.9^{+5.0}_{-3.7}$ & $19.6^{+3.5}_{-2.7}$ & $19.9^{+3.8}_{-2.9}$\\
~~~$\sigma_\textnormal{GP, HARPS}$ & \ldots & \ldots & $33.2^{+6.9}_{-5.4}$ & $25.5^{+5.6}_{-4.2}$ & $26.0^{+5.8}_{-4.3}$ & $25.9^{+5.9}_{-4.3}$ & $20.6^{+4.7}_{-3.4}$ & $21.2^{+4.7}_{-3.6}$ \\
~~~$\sigma_\textnormal{GP, CARMENES\ VIS1}$ & \ldots & \ldots & $26.6^{+9.8}_{-6.6}$ & $15.8^{+7.2}_{-4.8}$ & $16.9^{+7.4}_{-5.1}$ & $19.4^{+6.9}_{-4.8}$ & $13.1^{+5.4}_{-3.6}$ & $13.5^{+5.3}_{-3.8}$ \\
~~~$\sigma_\textnormal{GP, CARMENES\ NIR1}$ & \ldots & \ldots & $32^{+17}_{-13}$ & $0.78^{+13.00}_{-0.74}$ & $4.2^{+25.0}_{-4.1}$ & $18.5^{+23.0}_{-9.6}$ & $0.26^{+2.40}_{-0.23}$ & $1.4^{+9.4}_{-1.3}$ \\
~~~$\sigma_\textnormal{GP, HARPS-N1}$ & \ldots & \ldots & $26.0^{+8.7}_{-5.6}$ & $16.2^{+6.3}_{-4.1}$ & $17.3^{+6.3}_{-4.5}$ & $21.0^{+5.8}_{-4.3}$ & $14.6^{+4.4}_{-3.3}$ & $15.1^{+4.8}_{-3.4}$ \\
~~~$\sigma_\textnormal{GP, HARPS-N2}$ & \ldots & \ldots & $15.8^{+7.7}_{-5.0}$ & $10.3^{+7.3}_{-9.7}$ & $7.5^{+8.3}_{-7.3}$ & $30.2^{+10.0}_{-8.1}$ & $21.6^{+7.9}_{-7.9}$ & $20.7^{+8.5}_{-19.0}$ \\
~~~$\sigma_\textnormal{GP, CARMENES\ VIS2}$ & \ldots & \ldots & $17.8^{+8.5}_{-5.1}$ & $1.9^{+6.5}_{-1.9}$ & $5.7^{+5.8}_{-5.4}$ & $11.0^{+5.8}_{-3.3}$ & $0.38^{+2.90}_{-0.35}$ & $3.2^{+4.6}_{-3.0}$ \\
~~~$\sigma_\textnormal{GP, CARMENES\ NIR2}$ & \ldots & \ldots & $15.7^{+14.0}_{-8.3}$ & $7.1^{+11.0}_{-6.9}$ & $2.3^{+12.0}_{-2.3}$ & $9.3^{+12.0}_{-5.2}$ & $6.8^{+9.7}_{-5.7}$ & $2.4^{+9.2}_{-2.3}$ \\
~~~$\sigma_\textnormal{GP, HPF1}$ & \ldots & \ldots & $30^{+15}_{-12}$ & $17^{+14}_{-16}$ & $19^{+13}_{-18}$ & $23^{+16}_{-14}$ & $4.8^{+16.0}_{-4.7}$ & $13^{+12}_{-12}$ \\
~~~$\sigma_\textnormal{GP, HPF2}$ & \ldots & \ldots & $8.4^{+2.6}_{-2.0}$ & $9.2^{+2.7}_{-2.5}$ & $8.2^{+3.0}_{-2.7}$ & $6.1^{+2.7}_{-1.8}$ & $7.8^{+2.6}_{-1.8}$ & $7.2^{+2.5}_{-1.9}$ \\
~~~$\sigma_\textnormal{HIRES}$ & $17.0^{+2.0}_{-1.7}$ & $19.5^{+2.5}_{-2.1}$ & $1.4^{+1.2}_{-1.3}$ & $1.3^{+1.2}_{-1.2}$ & $1.3^{+1.1}_{-1.2}$ & $0.78^{+1.20}_{-0.73}$ & $0.69^{+1.20}_{-0.64}$ & $0.59^{+1.30}_{-0.55}$ \\
~~~$\sigma_\textnormal{HARPS}$ & $18.0^{+2.1}_{-1.7}$ & $12.2^{+1.5}_{-1.2}$ & $1.05^{+0.44}_{-0.46}$ & $0.93^{+0.48}_{-0.67}$ & $0.92^{+0.47}_{-0.66}$ & $0.12^{+0.48}_{-0.10}$ & $0.108^{+0.410}_{-0.086}$ & $0.094^{+0.370}_{-0.073}$\\
~~~$\sigma_\textnormal{CARMENES\ VIS1}$ & $7.4^{+1.4}_{-1.1}$ & $2.33^{+0.84}_{-0.82}$ & $0.48^{+1.40}_{-0.44}$ & $0.59^{+1.50}_{-0.55}$ & $0.61^{+1.40}_{-0.57}$ & $0.24^{+1.10}_{-0.21}$ & $0.26^{+1.10}_{-0.23}$ & $0.35^{+1.10}_{-0.31}$ \\
~~~$\sigma_\textnormal{CARMENES\ NIR1}$& $1.2^{+5.3}_{-1.2}$ & $11.6^{+3.2}_{-2.7}$ & $0.60^{+4.60}_{-0.56}$ & $1.4^{+5.8}_{-1.4}$ & $1.4^{+6.1}_{-1.4}$ & $0.89^{+6.00}_{-0.85}$ & $1.8^{+5.5}_{-1.7}$ & $2.0^{+5.9}_{-1.9}$ \\
~~~$\sigma_\textnormal{HARPS-N1}$ & $11.1^{+1.4}_{-1.2}$ & $4.05^{+0.81}_{-0.62}$ & $1.58^{+0.37}_{-0.32}$ & $1.62^{+0.37}_{-0.33}$ & $1.62^{+0.37}_{-0.32}$ & $0.40^{+0.62}_{-0.36}$ & $0.46^{+0.62}_{-0.42}$ & $0.51^{+0.58}_{-0.45}$ \\
~~~$\sigma_\textnormal{HARPS-N2}$ & $9.1^{+1.7}_{-1.3}$ & $16.6^{+3.1}_{-2.4}$ & $4.2^{+1.4}_{-1.1}$ & $4.9^{+1.9}_{-1.5}$ & $5.1^{+1.6}_{-1.6}$ & $0.18^{+1.20}_{-0.15}$ & $0.25^{+2.90}_{-0.22}$ & $0.42^{+5.00}_{-0.38}$ \\
~~~$\sigma_\textnormal{CARMENES\ VIS2}$ & $0.098^{+0.410}_{-0.077}$ & $8.9^{+1.9}_{-1.5}$ & $0.095^{+0.380}_{-0.075}$ & $0.098^{+0.380}_{-0.077}$ & $0.10^{+0.40}_{-0.08}$ & $0.097^{+0.380}_{-0.076}$ & $0.092^{+0.370}_{-0.071}$ & $0.110^{+0.360}_{-0.086}$ \\
~~~$\sigma_\textnormal{CARMENES\ NIR2}$ & $5.6^{+4.5}_{-5.3}$ & $6.1^{+4.0}_{-5.8}$ & $0.3^{+2.8}_{-0.3}$ & $0.28^{+2.50}_{-0.25}$ & $0.27^{+2.30}_{-0.24}$ & $0.3^{+3.0}_{-0.3}$ & $0.29^{+2.60}_{-0.26}$ & $0.36^{+2.80}_{-0.32}$ \\
~~~$\sigma_\textnormal{HPF1}$ & $10.2^{+5.1}_{-2.9}$ & $7.6^{+4.1}_{-2.2}$ & $0.89^{+12.00}_{-0.84}$ & $2.2^{+10.0}_{-2.1}$ & $1.4^{+10.0}_{-1.3}$ & $1.7^{+14.0}_{-1.6}$ & $7.9^{+6.0}_{-7.5}$ & $3.9^{+8.4}_{-3.7}$ \\
~~~$\sigma_\textnormal{HPF2}$ & $12.1^{+1.9}_{-1.5}$ & $20.7^{+3.0}_{-2.5}$ & $2.98^{+0.82}_{-0.65}$ & $2.50^{+0.71}_{-0.62}$ & $2.61^{+0.75}_{-0.62}$ & $2.90^{+0.72}_{-0.63}$ & $2.39^{+0.72}_{-0.74}$ & $2.47^{+0.73}_{-0.69}$ \\
\noalign{\smallskip}
\hline
\noalign{\smallskip}
~~~$\ln \mathcal{Z}$ & --1143.31 & --1143.38 & --1002.17 & --992.50 & --993.37 & --966.77 & \textbf{-963.40} & --965.73\\
~~~$\Delta\ln \mathcal{Z}$ & --179.91 & --179.99 & --38.77 & --29.10 & --29.98 & --3.37 & \textbf{0.0} & --2.33\\ 
\noalign{\smallskip}
\hline
\noalign{\smallskip}
   \end{tabular}
    \tablefoot{
        \tablefoottext{a}{Upper limits denote the 95\% upper credibility interval of fits.}
        \tablefoottext{b}{Priors for the NIR arm are slightly different because of the lower data quality. The period priors are wider, $\mathcal{U}$(2.2\,d, 2.5\,d).}
    }
\end{table*}

\section{Discussion and future outlook} \label{sec:discussion}
In this analysis, we presented new CARMENES optical and near-IR data that cover a wide wavelength range well into the red part of the spectrum to test the origin of the 2.23\,d signal found in the RVs for AD~Leo. The presence of stellar activity is supported with CARMENES through the proof of wavelength- and time-dependence. 
With the CARMENES data alone, the shape of the wavelength dependence of the RV amplitude can be attributed to a star-spot configuration following \cite{Reiners2010}. 
Between the two CARMENES seasons, the amplitude decreased in both instruments. Specifically, from $28.43^{+0.94}_{-0.98}$\,\ms\ to $17.02^{+0.63}_{-0.62}$\,\ms\ in the VIS channel, and from $18.0^{+2.6}_{-2.7}$\,\ms\ to $9.2^{+3.0}_{-2.6}$\,\ms\ in the near-IR channel when modeling the signals as pure sinsuoidal variations.
\citetalias{Carleo2020_ADLeo_GIARPS} also found an amplitude decrease between the HARPS-N and GIANO-B data, that is, from 33\,\ms\ to less than 23\,\ms\ between the instruments and from 33\,\ms\ to 13\,\ms\ between seasons. However, given the large errorbars of the GIANO-B data ($\sim$20\,\ms), it was only possible to identify an amplitude decrease, but not to constrain whether there even is a signal present in these data at 2.23\,d. 
Likewise, the high-precision near-IR data from HPF show a much smaller RV amplitude \citepalias{Robertson2020_ADLeo_HPF}, inconsistent with the higher RV amplitude in the optical, and there is also a discrepancy even between HPF seasons, namely, the RMS scatter dropped from 23\,\ms\ to 6.4\,\ms. This decrease in amplitude is also seen in the overlapping, optical HARPS-N data and second season of CARMENES data (see again Fig.~\ref{fig:allrv_timeseries}).
Nonetheless, both our results and those in \citetalias{Carleo2020_ADLeo_GIARPS} and \citetalias{Robertson2020_ADLeo_HPF} agree that the signal's amplitude does indeed decrease rather than increase, indicating that AD~Leo is entering a lower-activity phase. We conclude that the effect of the spots on the RVs is dominated by temperature differences rather than the Zeeman effect \citep[see][]{Reiners2013}, in accordance with the other evidence at hand such as photometric variability (presented in \citetalias{Tuomi2018_ADLeo}).

\subsection{Possibility of stable stellar-spot signal}

It is common practice to model activity-induced RVs with GPs \citep[e.g.,][and further citations]{Rajpaul2015}, as they are nondeterministic models that can sufficiently fit stochastic modulations. Provided that the evidence against a planet outweighs that in favor of a planet, our understanding was to expect that the GP-only model should have been adequate enough in accounting for the stellar activity, with the assumption that the signal in AD~Leo is purely stellar activity induced. The modeling instead showed that neither a stable-only (i.e., sinusoid) model nor red-noise-only (i.e., GP-only) model can correctly describe the whole data set (Table~\ref{tab:models}). A mix of a stable and variable component is rather the preferred model, where the stable model has an amplitude of ${K}_\textnormal{stable} = 16.6 \pm 2.2$\,\ms, lower than the value of 19\,\ms\ proposed by \citetalias{Tuomi2018_ADLeo}. 
We do not claim that the stable component is solely due to a planetary companion as it could perhaps be also due to the persistent presence of a circumpolar stellar spot imposing a constant behavior, or even a mixture of both. 
Thus, we set an 3$\sigma$ upper limit (i.e., from ${K}_\textnormal{stable} = 16.6 \pm 3 \times 2.2$\,\ms) of 27\,$\textnormal{M}_\oplus$ (or 0.084\,$\textnormal{M}_\textnormal{Jup}$) for a putative planetary companion, in comparison to the proposed mass of $\sim$0.2\,$\textnormal{M}_\textnormal{Jup}$ by \citetalias{Tuomi2018_ADLeo}.
Focusing on the amplitudes of the GP kernel of the mixed model, we found that those of the earlier optical RV data, specifically HARPS and HIRES ($\sigma_\textnormal{GP,\ inst} \sim 20$\,\ms), were  higher than most of the more recent optical data ($\sigma_\textnormal{GP,\ inst} \lesssim 13$\,\ms). These values support the assumption that the star was more active earlier and is entering a less active phase (Sect.~\ref{sec:phot}). The only exception was with HT2, with $\sigma_\textnormal{GP,\ HT2} \sim 22$\,\ms, an even higher value. The GP amplitudes of near-IR spectrographs were consistent with $\sigma_\textnormal{GP,\ inst} \sim$ 0\,\ms, except for HPF2 ($\sigma_\textnormal{GP,\ HPF2} \sim 8$\,\ms). The data from the two instruments that stand out (i.e., HT2 and HPF2) were contemporaneously taken, indicating that there was some phase shift between the stable and variable component during this time period. A better understanding of how model comparison can behave given a purely stellar-activity induced signal would be necessary to determine if such a result is even expected (see also Sect.~\ref{sec:limitations} for future suggestions). 
Additionally, our analysis of recomputing the RVs, considering only those spectral lines unaffected by stellar activity, still showed some strong significant periodicity at 2.23\,d (Fig.~\ref{fig:spectrallines}), unlike for other known active stars where the signals had disappeared (Lafarga priv. comm.).

Nonetheless, this should raise the question of whether a signal with an amplitude of $\sim$17\,\ms\ that is fully stable over all these observations ($\sim$19\,yr) in such an M dwarf is possible. In fact, it is not so surprising that spot-induced RV fluctuations for M dwarfs are long-lived \citep[e.g.,][]{Guenther2020,Quirrenbach2022}. 
Though evidence of stable stellar activity behavior had previously been shown in photometry \citep[e.g., GJ~1243,][]{Davenport2020} and RVs \citep[e.g., $\alpha$~Tau,][]{Hatzes2015} over time, our paper demonstrates the first case for modeling RVs over time and wavelength. 
Future studies performing simulations with software packages such as \texttt{StarSim 2.0} \citep{Herrero2016_starsim} or \texttt{SOAP 2.0} \citep{Dumusque2014} using various star-spot configurations may also shed light on why a signal can stay persistent over many years \citep[][]{Herrero2016_starsim,Rosich2020}. Utilizing \texttt{StarSim 2.0} to compare RV-CRX correlations between simulated and real data as performed by \cite{Baroch2020}, with YZ~CMi as a case study, could be beneficial in determining the star-spot temperature difference, the star-spot filling factor, and the location of the spot. A first test with a simple assumption (i.e., one big polar spot) could reproduce well the RVs and CRX values of AD~Leo for the CARMENES VIS data (Baroch priv. comm.). But such an approach can become degenerate when considering so many instruments and various star-spot configurations, and a planetary signal would act as an achromatic offset to the RVs and CRX.

\subsection{Current limitations} \label{sec:limitations}

AD~Leo is a prime case study for M dwarfs to explore the possibility of a planetary signal with an orbital period indistinguishably close to its stellar host's rotation period. Given its rich multi-wavelength spectroscopic monitoring over a large time baseline, there are no other targets with such an extensive data set coverage. Even then, we showed that there are apparent holes in our current state of research in this field in which studies to address them would require proposals, dedicated surveys and telescope time. 
In essence, we are either limited by our modeling approaches, by our astrophysical knowledge of star-spot configurations, or a combination of both. In order to progress and disentangle this problem, we propose the following, in no particular order.

\subsubsection{Simultaneous multiband photometry}
The shape of the curve in Fig.~\ref{fig:Kvswavelength} is determined by stellar activity behavior, whereas the presence of a planet would simply act as an offset across the wavelength space. 
Even when doubling (or halving) the number of available spectroscopic data points, disentangling the contribution of a planetary signal within this curve still persists as a degenerate problem. 
Obtaining continuous and simultaneous multiband photometry would help in painting a better picture of the stellar-spot map distribution. In doing so, this can then be translated by forward modeling to determine the stellar-activity induced RVs \citep[e.g., via \texttt{Starsim 2.0} or \texttt{SOAP 2.0};][]{Herrero2016_starsim,Dumusque2014}. The residual between these computed RVs and those obtained would indicate the contribution of the Keplerian component. However, the caveat is, that spot configurations on the stellar surface can vary with time. Thus, continuous photometry would only apply to simultaneously taken RVs. 

\subsubsection{A better-suited RV data set for spectral line analyses}
In the absence of available photometric facilities, the approach of selecting spectral lines not affected by activity seems promising. This line-by-line approach has been successfully applied to other stars \citep[e.g.,][]{Dumusque2018_spectrallines,Wise2018activelines,Cretignier2020indivline}. In the case for AD~Leo, it is difficult to decipher whether it was the target itself or the relatively new method that resulted in an inconclusive outcome. Perhaps the RVs were not as affected by the stellar spots (i.e., due to the pole-on inclination), the RV amplitude was not as large as the others, or there were not enough data points to strengthen the correlation plots by establishing a statistically stronger correlation. Definitely, if overall activity decreases, then it would be more difficult to pick up the activity within certain spectral lines. Hence, observing a star when it is in its most active phase would constitute the ideal data set. Additionally, our spectral line analysis should be applied to a wider wavelength range (e.g., HARPS) as well to determine if the signal disappears. Another possible explanation could be that all lines are inherently affected by activity to some degree. Nonetheless, determining the ideal targets for this technique as well as developing the method further will be paramount in suppressing activity for future M-dwarf studies.

\subsubsection{Magnetic star-planet interaction} \label{sec:starplanetinteractoin}
The radio bursts thus far observed for the AD~Leo system (outlined in Sect.~\ref{sec:adleoradio}) suggest an excess that is unaccounted for. A possible scenario to justify the discrepancy could be due to sub-Alfv\'enic interaction of the stellar magnetosphere with an orbiting body, in this case with a putative planet around AD~Leo, which has $P_{\rm orb} = P_{\rm rot}$. 
This magnetic interaction between the star and a planet (Magnetic Star-Planet Interaction; MSPI) would be responsible for producing: (1) an enhancement of the magnetic activity of the star, and (2) the long-duration radio emission as that detailed in Sect.~\ref{sec:adleoradio}, both effects of which are modulated in phase with the orbital period of the planet.
In a star like AD~Leo, the difficulties for detecting this modulation are that this target is seen nearly pole-on ($i \sim$ 13\,deg) and that a putative planet orbit is synchronized with the stellar rotation. This produces a weaker MSPI, as it depends on the relative velocity of the planet with respect to the magnetosphere of the star \citep{Lanza2009,Lanza2012} and makes it difficult to disentangle it from stellar intrinsic activity. However, the presence of such a planet could explain the need for the stable component and part of its semi-amplitude ($K_\textnormal{stable}$ < 16.6\,\ms) seen in the RVs in this work. MSPI would show as a hot spot in the chromosphere that would mainly affect the chromospheric activity indicators such as $\log{R'_\textnormal{HK}}$ from the Ca~\textsc{ii}~H\&K lines, not available in the CARMENES data set but covered by HARPS.

Within this scenario, we followed the prescriptions in Appendix~B of \citet{PerezTorres2021} to estimate the flux density expected to arise from the interaction between a putative planet and its host star. We assume for simplicity an isothermal Parker wind with a temperature of (2--3)$\times 10^6$\,K, and a density of 10$^{7--8}$\,cm$^{-3}$ at the base of the corona. Since the hypothetical planet would be very close to its host star, we find that the planet is in the sub-Alfv\'enic regime, which makes it possible that energy and momentum can be carried upstream of the flow along Alfv\'en wings. We estimate that the radio flux in the few GHz band, arising from star-planet interaction, is in the range from less than about 0.1\,mJy and up to a few mJy, or even higher, depending on the stellar wind parameters at both the corona and at the orbital position of the planet. We therefore encourage a campaign for simultaneous monitoring of AD~Leo in both optical (i.e., spectroscopy, including Ca~\textsc{ii}~H\&K lines, and photometry) and in radio wavelengths to test the star-planet (sub-Alfv\'enic) interaction scenario. 
The detection of a correlation in the radio signal with the orbital period would be the smoking gun of sub-Alfv\'enic interaction, serving as an argument in support of the presence of a planet in the system.

\subsubsection{Application of GP models for red-noise correlated signals}

Additional data may not be what is necessary as much as is a better understanding of the astrophysical effects stellar spots impose on the star and how we choose to model them through nonphysically motivated means. 
The usage of GPs is currently the most popular approach in the RV field for modeling stellar activity. That being said, the most commonly used kernels (i.e., QP-GP, dSHO-GP) are not completely physically motivated, but nonetheless serve as a very good approximation to our aim of describing the quasi-periodic behavior. They can play a vital role, especially when trying to search for low-mass planets with low amplitudes among noisy data \citep[e.g.,][]{Stock2020_YZCeti,Demangeon2021,Faria2022}. 

The problem is, however, that we do not know just how effective the GPs are in serving their purpose and what limitations and consequences on (planetary) parameters they particularly have. Perhaps, the choice of the GP for different time baselines could be also of importance, as is the case for AD~Leo (see Sect.~\ref{sec:timedependence}), considering there are, for the time being, too many unknowns that can lead to misinterpretations. Newer implementations of GP kernels with a potentially better physical interpretation of stellar processes could become essential \citep[e.g.,][]{Perger2021,Luger2021}.
Applying these models to real data sets turns out to not be the solution, as it becomes complicated to correctly interpret the results. For this reason, these models should be compared against each other on controlled, simulated RV time series. 
A more in-depth study could include simulating RV time series by injecting solely stellar-activity-induced signals and a mixture of stellar activity and planetary components for a general grasp \citep{Stock2022}.
One step further would be to specifically simulate the stellar characteristics of AD~Leo, together with the time stamps and various wavelengths.
Even here, the limitation of our knowledge of spot distribution maps and correctly choosing the correct one persists. 

In future potential cases where there is a presumed planetary and stellar activity signal with two instruments covering a wide wavelength range, for instance, with CARMENES, it is crucial to adequately model the wavelength dependence of the stellar activity to ensure the most precise planetary parameters. More so for M-dwarf stars considering that their stellar rotation periods would often correlate to the orbital periods of planets residing in the habitable zone \citep{Newton2016}.
AD~Leo in this regard is an interesting case study to perform such fits when considering real data, however, simulated data sets may offer a more controlled environment to test the effectiveness of our current modeling techniques. 
Either way, AD~Leo certainly serves as a particularly intriguing system for studying the impact of a stable spot periodicity on the search for planetary signals.

\subsubsection{Using auxiliary data to correct for stellar activity} \label{sec:auxdata}
Another promising procedure is to include the stellar activity indicators as auxiliary data to try to detrend the RVs, commonly done so as linear components \citep[e.g.,][]{AngladaEscude2016_proxcentauri} or within the GP kernel \citep[e.g.,][]{SuarezMascareno2020,Cale2021_aumic,pyaneti2022} in regards to planet detection. Likewise, this technique is being carried out in a sole stellar activity context \citep{Jeffers2022,Cardona2022}, with particular attention to not only detrending with various indicators, but also to characterizing the behavior of the correlations. Specifically, this ``closed-loop'' relation is becoming more prevalent, as already exhibited with known active M-dwarf stars such as \object{YZ~CMi} \citep{Zechmeister2018_SERVAL,Baroch2020}, EV~Lac \citep{Jeffers2022}, and \object{GJ~674} \citep{Bonfils2007}, and now for AD~Leo as displayed in the RV-CRX and RV-BIS correlations (Fig.~\ref{fig:carmvisactivity}). Subtracting out the CRX dependency can improve the rms of the RVs, though it is nearly impossible to obtain a straight line from the correction, that is, completely removing all of the stellar activity, even in cases with very high signal-to-noise ratios and very dense sampling \citep[e.g., factor of three improvement for EV~Lac;][]{Jeffers2022}. For our case of AD~Leo (Sect.~\ref{sec:crx}), the CRX detrending improved the rms roughly by a factor of two for VIS1 (from 18.5\,\ms to 9.2\,\ms).

In this line, a relatively recent technique is that of using spectropolarimetric measurements to correct the RVs of the variability introduced by the activity.
The data are used to derived precise RVs and to obtain maps of the magnetic field and the distribution of surface inhomogeneities.
These maps are then used to obtain a model RV curve which are the subtracted from the observed RVs.
The removal of activity variations with this technique decreased the amplitude of the RV curve in V830\,Tau by a factor 10, which allowed the detection of a hot Jupiter orbiting this very young T\,Tauri star \citep{Donati2016}.

While these approaches are encouraging, they still poses limitations when considering a putative planetary signal whose orbital period is equal to the rotational period because the same issue, as mentioned already above, of degeneracy still holds true. Again, a systematic simulated analysis of forward modeling RVs and stellar activity indicators could be key in understanding what correlations to expect and how to properly detrend them.

\section{Conclusions} \label{sec:conclusions}
In this paper we presented new CARMENES optical and near-infrared spectroscopic data for a known nearby, active M dwarf, AD~Leo. The stellar rotation period of 2.23\,d is clearly present in the RVs, and we address the question of whether there could be a planet orbiting with the same periodicity, $P_{\rm orb} = P_{\rm rot}$. Taking advantage of the wide wavelength range of the CARMENES instrument, we demonstrated the shape of the wavelength dependency of the RV semi-amplitudes to be in agreement with what is expected for a star-spot temperature difference configuration. The strong anticorrelation found between CRX-RV and BIS-RV for the CARMENES VIS data were additional signs of stellar activity. In addition, we recomputed RVs using spectral lines unaffected by stellar activity, which has not been done before for AD~Leo, and found that there was still some significant residual power at the rotational period. 

When incorporating all available RVs (HARPS, HIRES, CARMENES, and the more recent HARPS-N, GIANO-B, and HPF data), it became evident that the signal has undergone various amplitude fluctuations as well as phase shifts, and this is behavior that cannot be attributed to the presence of a planetary companion. However, a closer look into the model comparison showed that a mixed model of a stable plus a quasi-periodic red-noise model best explained the data, where the stable component had a semi-amplitude of $K_\textnormal{stable}$ = $16.6 \pm 2.2$\,\ms, setting a 3$\sigma$ upper limit on the mass of a putative planet, $27\,\textnormal{M}_{\oplus}$ ($= 0.084\,\textnormal{M}_\textnormal{Jup}$). Based on all this evidence, our current machinery, and the given data at hand, we conclude we cannot unequivocally prove, or even disprove, the 2.23\,d periodic signal found in AD~Leo to be solely due to stellar activity. We therefore suggest to obtain simultaneous photometry and spectroscopy in various wavelengths accompanied by radio observations, and complemented with a clearer insight of stellar activity behavior on measurements via simulated time series to break the degeneracy.



\begin{acknowledgements}
CARMENES is an instrument for the Centro Astron\'omico Hispano-Alem\'an de Calar Alto (CAHA, Almer\'{\i}a, Spain).   
CARMENES was funded by the Max-Planck-Gesellschaft (MPG), 
 the Consejo Superior de Investigaciones Cient\'{\i}ficas (CSIC),
 the Ministerio de Econom\'ia y Competitividad (MINECO) and the European Regional Development Fund (ERDF) through projects FICTS-2011-02, ICTS-2017-07-CAHA-4, and CAHA16-CE-3978, 
 and the members of the CARMENES Consortium 
 (Max-Planck-Institut f\"ur Astronomie,
 Instituto de Astrof\'{\i}sica de Andaluc\'{\i}a,
 Landessternwarte K\"onigstuhl,
 Institut de Ci\`encies de l'Espai,
 Institut f\"ur Astrophysik G\"ottingen,
 Universidad Complutense de Madrid,
 Th\"uringer Landessternwarte Tautenburg,
 Instituto de Astrof\'{\i}sica de Canarias,
 Hamburger Sternwarte,
 Centro de Astrobiolog\'{\i}a and
 Centro Astron\'omico Hispano-Alem\'an), 
 with additional contributions by the MINECO, 
 the Deutsche Forschungsgemeinschaft (DFG) through the Major Research Instrumentation Programme and Research Unit FOR2544 ``Blue Planets around Red Stars'', 
 the Klaus Tschira Stiftung, 
 the states of Baden-W\"urttemberg and Niedersachsen, 
 and by the Junta de Andaluc\'{\i}a.
We acknowledge financial support from the Agencia Estatal de Investigaci\'on of the Ministerio de Ciencia, Innovaci\'on y Universidades and the ERDF through projects 
PID2019-109522GB-C51/2/3/4	
 PGC2018-098153-B-C33		
 AYA2017-84089, 		
 AYA2016-79425-C3-1/2/3-P,	
 ESP2016-80435-C2-1-R,		
 and the Centre of Excellence ``Severo Ochoa'' and ``Mar\'ia de Maeztu'' awards to the Instituto de Astrof\'isica de Canarias (SEV-2015-0548), Instituto de Astrof\'isica de Andaluc\'ia (SEV-2017-0709), and Centro de Astrobiolog\'ia (MDM-2017-0737), the Generalitat de Catalunya/CERCA programme, and the DFG under Research Unit FOR2544. 
 T.T. acknowledges support by the DFG Research Unit FOR 2544 "Blue
Planets around Red Stars" project No. KU 3625/2-1.
T.T. further acknowledges support by the BNSF program ``VIHREN-2021''
project No.~KP-06-DV/5.

This research has made use of the ESO Science Archive Facility and the CARMENES data archive at CAB (CSIC-INTA). 
Based on observations collected at the European Southern Observatory under ESO programs 072.C-0488, 191.C-0505, and 192.C-0224. 
\end{acknowledgements}

\bibliographystyle{aa} 
\bibliography{biblio} 

\newpage
\begin{appendix} 
\clearpage

 \section{RV data and short tables} \label{appendix:rvdata}
 
\begin{table}[bp]
\centering 
\small
\begin{center}
\caption{Published rotational periods values of AD~Leo\tablefootmark{a}.}
\label{tab:protvalues}
\centering
\begin{tabular}{lcl}
\hline \hline
\noalign{\smallskip}
Data & $P_\textnormal{rot}$ & Reference \\
 & (d) &  \\
\noalign{\smallskip}
\hline
\noalign{\smallskip}
\multicolumn{3}{c}{\textit{Photometry}}\\
\noalign{\smallskip}
TESS-s48  & $2.2304 \pm 0.0014$ & This work\tablefootmark{b} \\
STELLA & $2.237 \pm 0.035$ & \citetalias{Carleo2020_ADLeo_GIARPS}\\
ASAS-N & $2.22791^{+0.00066}_{-0.00055}$ & \citetalias{Tuomi2018_ADLeo}\tablefootmark{c}\\
\textit{MOST}-1 & $2.289 \pm 0.019$ & \citetalias{Tuomi2018_ADLeo}\tablefootmark{d}\\
\textit{MOST}-2 & $2.145 \pm 0.011$ & \citetalias{Tuomi2018_ADLeo}\tablefootmark{d}\\
\textit{MOST} & $2.23^{+0.36}_{-0.27}$ & Hun2012\\
FCAPT & 2.23& Eng2009\\
\noalign{\smallskip}
\multicolumn{3}{c}{\textit{Spectroscopy}}\\
\noalign{\smallskip}
HP2 & 2.2249 & \citetalias{Robertson2020_ADLeo_HPF}\tablefootmark{e}\\
HT1 & $2.2244 \pm 0.0010$ & \citetalias{Carleo2020_ADLeo_GIARPS}\\
HT2 & $2.2225 \pm 0.0044$ & \citetalias{Carleo2020_ADLeo_GIARPS}\\
G1 + G2 & $2.2246 \pm 0.0219$ & \citetalias{Carleo2020_ADLeo_GIARPS}\tablefootmark{f}\\
HARPS + HIRES & $2.22567^{+0.00026}_{-0.00011}$ & \citetalias{Tuomi2018_ADLeo}\\
HARPS & 2.22704 & Rei2013\tablefootmark{g} \\ 
HARPS & $2.2267 \pm 0.0001$ & Bon2013 \\ 
\noalign{\smallskip}
\multicolumn{3}{c}{\textit{Tomography}}\\
\noalign{\smallskip}
ESPaDOnS + NARVAL  & $2.2399 \pm 0.0006$ & Mor2008 \\
\noalign{\smallskip}
\hline
\end{tabular}
\tablefoot{
\tablefoottext{a}{The $P_\textnormal{rot}$ determined by us from the GP component of the best-fit model on the spectroscopic data (Sect~\ref{sec:modelresults}, Table~\ref{tab:models}) is $2.2270^{+0.0010}_{-0.0011}$\,d.}
\tablefoottext{b}{See Sect.~\ref{sec:phot} for the analysis setup.}
\tablefoottext{c}{After subtraction of a longer periodicity in the data, see reference for details.}
\tablefoottext{d}{The \textit{MOST} baseline of 8.9\,d was separated into two independent data sets, see reference for details.}
\tablefoottext{e}{Value was not explicitly included in the text but appeared in the RV phase-folded plot.}
\tablefoottext{f}{A narrowly constrained Gaussian prior of $2.225 \pm 0.020$\,d was used to detect a 3$\sigma$ upper limit of $K = 23$\,\ms.}
\tablefoottext{g}{Value based off of the most prominent peak in the periodogram.}

}
\tablebib{
    Bon2013: \cite{Bonfils2013};
    Eng2009: \cite{Engel2009};
    Hun2012: \cite{HuntWalker2012};
    Mor2008: \cite{Morin2008};
    Rei2013: \cite{Reiners2013}.
} 

\end{center}
\end{table}

\begin{table}[!h]
\centering
\caption{CARMENES VIS RV data of AD~Leo.}
\label{tab:carmvisdata}
\begin{tabular}{c
                S[table-format=3]
                S[table-format=2]
                }
\hline
\hline
\noalign{\smallskip}
BJD (TDB\tablefootmark{*}) & \multicolumn{1}{c}{RV (m\,s$^{-1}$)} & \multicolumn{1}{c}{$\sigma_\textnormal{RV}$ (m\,s$^{-1}$)}  \\
\noalign{\smallskip}
\hline
\noalign{\smallskip}
2458199.31070 & 17.26 & 3.18\\[0.1 cm]
2458199.43446 & -0.85 & 2.68\\[0.1 cm]
2458199.53410 & -7.02 & 2.70\\[0.1 cm]
2458200.30763 & -13.30 & 1.61\\[0.1 cm]
2458200.33569 & -12.72 & 1.91\\[0.1 cm]
2458200.52737 & -0.81 & 2.85\\[0.1 cm]
2458205.33771 & 18.22 & 1.62\\[0.1 cm]
2458205.43139 & 23.51 & 1.59\\[0.1 cm]
2458205.57006 & 24.40 & 2.42\\[0.1 cm]
2458209.35609 & -6.66 & 1.37\\[0.1 cm]
2458209.47549 & 3.29 & 1.75\\[0.1 cm]
2458210.31450 & 21.81 & 7.25\\[0.1 cm]
2458211.33225 & -15.82 & 4.55\\[0.1 cm]
2458211.33586 & -12.85 & 3.16\\[0.1 cm]
2458212.36090 & 32.09 & 2.45\\[0.1 cm]
2458212.56685 & 25.08 & 2.78\\[0.1 cm]
2458213.41097 & -23.28 & 2.34\\[0.1 cm]
2458213.51416 & -22.32 & 2.31\\[0.1 cm]
2458215.31521 & -22.16 & 2.03\\[0.1 cm]
2458215.44196 & -31.14 & 2.08\\[0.1 cm]
2458215.54733 & -25.31 & 2.17\\[0.1 cm]
2458216.36673 & 5.91 & 3.27\\[0.1 cm]
2458217.35751 & -7.13 & 3.32\\[0.1 cm]
2458218.35496 & -8.17 & 4.46\\[0.1 cm]
2458225.34239 & 9.43 & 1.61\\[0.1 cm]
2458225.57485 & 24.92 & 2.99\\[0.1 cm]
\noalign{\smallskip}
\hline
\noalign{\medskip}
2458893.39518 & 21.10 & 2.58\\[0.1 cm]
2458893.39739 & 18.67 & 2.96\\[0.1 cm]
2458893.56436 & 18.29 & 2.49\\[0.1 cm]
2458893.56616 & 16.85 & 2.35\\[0.1 cm]
2458893.69412 & 14.18 & 2.18\\[0.1 cm]
2458893.69600 & 14.00 & 1.91\\[0.1 cm]
2458894.39429 & -14.04 & 2.12\\[0.1 cm]
2458894.39632 & -9.74 & 2.55\\[0.1 cm]
2458894.40312 & -12.89 & 2.22\\[0.1 cm]
2458894.40435 & -12.00 & 2.12\\[0.1 cm]
2458894.55514 & -15.42 & 1.90\\[0.1 cm]
2458894.55639 & -17.37 & 1.86\\[0.1 cm]
2458894.70318 & -16.27 & 2.50\\[0.1 cm]
2458894.70436 & -15.28 & 2.15\\[0.1 cm]
2458895.38983 & 11.95 & 2.90\\[0.1 cm]
2458895.39120 & 7.18 & 2.46\\[0.1 cm]
2458895.54067 & 14.16 & 2.47\\[0.1 cm]
2458895.54207 & 16.96 & 2.62\\[0.1 cm]
2458895.71912 & 21.55 & 3.84\\[0.1 cm]
2458895.72044 & 16.96 & 3.15\\[0.1 cm]
\noalign{\smallskip}
\hline
\end{tabular}
\tablefoot{\tablefoottext{*}{Barycentric dynamical time.}}
\end{table}

\begin{table}[!h]
\centering
\caption{CARMENES NIR RV data of AD~Leo.}
\label{tab:carmnirdata}
\begin{tabular}{c
                S[table-format=3]
                S[table-format=2]
                }
\hline
\hline
\noalign{\smallskip}
BJD (TDB\tablefootmark{*}) & \multicolumn{1}{c}{RV (m\,s$^{-1}$)} & \multicolumn{1}{c}{$\sigma_\textnormal{RV}$ (m\,s$^{-1}$)}  \\
\noalign{\smallskip}
\hline
\noalign{\smallskip}
2458199.31074 & -9.74 & 9.77\\[0.1 cm]
2458199.43432 & -1.48 & 9.80\\[0.1 cm]
2458199.53395 & -21.53 & 7.61\\[0.1 cm]
2458200.30763 & -16.58 & 5.70\\[0.1 cm]
2458200.33556 & -24.46 & 6.47\\[0.1 cm]
2458200.52732 & 5.43 & 24.79\\[0.1 cm]
2458205.33748 & 13.42 & 6.57\\[0.1 cm]
2458205.43117 & 27.81 & 5.84\\[0.1 cm]
2458205.57007 & 9.75 & 5.85\\[0.1 cm]
2458209.35584 & -5.05 & 5.92\\[0.1 cm]
2458209.47579 & -11.00 & 7.33\\[0.1 cm]
2458210.31440 & -5.64 & 23.89\\[0.1 cm]
2458211.33294 & -24.69 & 13.57\\[0.1 cm]
2458211.33596 & -21.72 & 9.17\\[0.1 cm]
2458212.36111 & 26.48 & 5.11\\[0.1 cm]
2458212.56689 & 7.25 & 7.70\\[0.1 cm]
2458213.41095 & -14.93 & 6.10\\[0.1 cm]
2458213.51413 & -16.77 & 5.91\\[0.1 cm]
2458215.31521 & -7.97 & 4.84\\[0.1 cm]
2458215.44212 & -6.86 & 5.62\\[0.1 cm]
2458215.54747 & -11.64 & 7.33\\[0.1 cm]
2458216.36658 & -7.24 & 9.65\\[0.1 cm]
2458217.35744 & 6.59 & 23.42\\[0.1 cm]
2458218.35494 & -9.44 & 11.83\\[0.1 cm]
2458225.34234 & 10.32 & 7.20\\[0.1 cm]
2458225.57469 & -1.85 & 10.24\\[0.1 cm]
\noalign{\smallskip}
\hline
\noalign{\medskip}
2458893.39513 & 17.18 & 5.81\\[0.1 cm]
2458893.39739 & 31.12 & 12.16\\[0.1 cm]
2458893.56431 & 8.59 & 7.67\\[0.1 cm]
2458893.56616 & 6.39 & 10.36\\[0.1 cm]
2458893.69395 & -2.56 & 11.05\\[0.1 cm]
2458893.69590 & 15.99 & 13.10\\[0.1 cm]
2458894.39425 & -1.52 & 7.95\\[0.1 cm]
2458894.39614 & -3.07 & 12.02\\[0.1 cm]
2458894.40311 & 15.02 & 8.90\\[0.1 cm]
2458894.40428 & 0.83 & 6.93\\[0.1 cm]
2458894.55510 & -2.68 & 7.32\\[0.1 cm]
2458894.55633 & -13.38 & 7.02\\[0.1 cm]
2458894.70320 & -27.15 & 11.60\\[0.1 cm]
2458894.70442 & -15.44 & 39.11\\[0.1 cm]
2458895.38980 & 18.25 & 6.22\\[0.1 cm]
2458895.39117 & 2.50 & 8.46\\[0.1 cm]
2458895.54064 & 10.29 & 7.53\\[0.1 cm]
2458895.54203 & -1.03 & 10.48\\[0.1 cm]
2458895.71911 & 13.87 & 14.45\\[0.1 cm]
2458895.72021 & 5.28 & 18.57\\[0.1 cm]
\noalign{\smallskip}
\hline
\end{tabular}
\tablefoot{\tablefoottext{*}{Barycentric dynamical time.}}
\end{table}

\begin{table}[!h]
\centering
\caption{HARPS RV and accompanying data for AD~Leo used in this paper, first processed using \serval\ \citep{Zechmeister2018_SERVAL} and then corrected for nightly zero points \citep{Trifonov2020}.}
\label{tab:harpsdata}
\begin{tabular}{c
                S[table-format=3]
                S[table-format=2]
                }
\hline
\hline
\noalign{\smallskip}
BJD (TDB\tablefootmark{*}) & \multicolumn{1}{c}{RV (m\,s$^{-1}$)} & \multicolumn{1}{c}{$\sigma_\textnormal{RV}$ (m\,s$^{-1}$)}  \\
\noalign{\smallskip}
\hline
\noalign{\smallskip}
2452986.859 & 28.07 & 1.89\\[0.1 cm]
2453511.548 & -19.61 & 1.01\\[0.1 cm]
2453520.521 & -15.62 & 1.00\\[0.1 cm]
2453543.482 & 13.69 & 1.26\\[0.1 cm]
2453544.453 & -9.50 & 0.92\\[0.1 cm]
2453550.460 & 39.00 & 2.32\\[0.1 cm]
2453728.866 & 39.05 & 1.03\\[0.1 cm]
2453758.755 & -21.83 & 0.74\\[0.1 cm]
2453760.755 & -12.09 & 0.76\\[0.1 cm]
2453761.781 & 18.34 & 0.67\\[0.1 cm]
2453783.726 & -7.40 & 0.70\\[0.1 cm]
2453785.727 & -18.55 & 0.69\\[0.1 cm]
2453809.661 & -2.50 & 0.56\\[0.1 cm]
2453810.677 & 10.93 & 0.68\\[0.1 cm]
2453811.676 & 8.04 & 0.66\\[0.1 cm]
2453812.664 & -9.53 & 0.96\\[0.1 cm]
2453813.659 & 20.68 & 0.66\\[0.1 cm]
2453814.655 & -20.46 & 0.85\\[0.1 cm]
2453815.571 & 26.14 & 0.72\\[0.1 cm]
2453815.622 & 25.40 & 0.66\\[0.1 cm]
2453815.736 & 24.43 & 0.63\\[0.1 cm]
2453816.544 & -14.43 & 0.74\\[0.1 cm]
2453816.656 & -19.40 & 0.65\\[0.1 cm]
2453816.722 & -21.99 & 0.68\\[0.1 cm]
2453817.551 & 24.31 & 0.63\\[0.1 cm]
2453817.676 & 26.22 & 0.67\\[0.1 cm]
2453829.615 & -1.07 & 0.62\\[0.1 cm]
2453830.540 & -7.15 & 0.79\\[0.1 cm]
2453831.670 & 8.61 & 0.57\\[0.1 cm]
2453832.650 & -11.77 & 0.62\\[0.1 cm]
2453833.628 & 23.40 & 0.57\\[0.1 cm]
2453834.611 & -27.06 & 0.63\\[0.1 cm]
2453835.656 & 27.56 & 0.63\\[0.1 cm]
2453836.617 & -19.54 & 0.59\\[0.1 cm]
2453861.594 & -1.22 & 0.69\\[0.1 cm]
2453863.569 & -22.08 & 0.69\\[0.1 cm]
2453864.535 & 29.03 & 0.66\\[0.1 cm]
2453867.542 & -13.93 & 0.69\\[0.1 cm]
2453868.518 & 22.20 & 0.70\\[0.1 cm]
2453871.563 & 19.01 & 0.62\\[0.1 cm]
2456656.850 & -36.79 & 0.87\\[0.1 cm]
2456656.861 & -35.19 & 0.86\\[0.1 cm]
2456657.853 & 39.91 & 0.88\\[0.1 cm]
2456658.865 & -45.31 & 0.88\\[0.1 cm]
2456658.876 & -45.60 & 0.86\\[0.1 cm]
2456659.859 & 43.23 & 1.17\\[0.1 cm]
2456797.513 & -29.24 & 0.57\\[0.1 cm]
\noalign{\smallskip}
\hline
\end{tabular}
\tablefoot{\tablefoottext{*}{Barycentric dynamical time.}}
\end{table}

\begin{table}[!h]
\centering
\caption{HIRES RV data for AD~Leo used in this paper produced by \cite{Tal-Or2019_HIRES}.}
\label{tab:hiresdata}
\begin{tabular}{c
                S[table-format=3]
                S[table-format=2]
                }
\hline
\hline
\noalign{\smallskip}
BJD (TDB\tablefootmark{*}) & \multicolumn{1}{c}{RV (m\,s$^{-1}$)} & \multicolumn{1}{c}{$\sigma_\textnormal{RV}$ (m\,s$^{-1}$)}  \\
\noalign{\smallskip}
\hline
\noalign{\smallskip}
2452064.85709 & -4.47 & 2.24\\[0.1 cm]
2452334.01660 & 28.75 & 1.89\\[0.1 cm]
2452602.13013 & -17.43 & 1.94\\[0.1 cm]
2452652.10228 & 8.91 & 1.97\\[0.1 cm]
2452653.01007 & -5.16 & 1.98\\[0.1 cm]
2452711.93243 & -4.57 & 1.80\\[0.1 cm]
2452712.95569 & -5.29 & 2.08\\[0.1 cm]
2452804.84095 & -14.89 & 2.40\\[0.1 cm]
2452828.76276 & 5.11 & 2.12\\[0.1 cm]
2453044.88295 & 17.33 & 2.10\\[0.1 cm]
2453339.12059 & 13.73 & 2.02\\[0.1 cm]
2453340.13691 & -24.87 & 2.03\\[0.1 cm]
2453398.91738 & 16.05 & 1.94\\[0.1 cm]
2453398.92487 & 12.70 & 1.82\\[0.1 cm]
2453724.00531 & 22.43 & 2.11\\[0.1 cm]
2453724.01159 & 20.59 & 1.88\\[0.1 cm]
2453748.00269 & -21.62 & 1.73\\[0.1 cm]
2453749.89640 & -21.39 & 1.72\\[0.1 cm]
2453749.90307 & -24.10 & 1.83\\[0.1 cm]
2453750.91326 & 15.30 & 1.55\\[0.1 cm]
2453751.93398 & -12.65 & 3.19\\[0.1 cm]
2453753.00706 & 17.96 & 1.75\\[0.1 cm]
2453753.01294 & 28.22 & 2.16\\[0.1 cm]
2453753.94933 & -12.15 & 1.71\\[0.1 cm]
2453753.95561 & -9.32 & 1.73\\[0.1 cm]
2453776.07390 & -0.55 & 1.69\\[0.1 cm]
2453776.08012 & -0.77 & 1.59\\[0.1 cm]
2453777.06086 & -22.01 & 1.65\\[0.1 cm]
2453777.06716 & -17.44 & 1.73\\[0.1 cm]
2453777.88577 & 8.69 & 1.56\\[0.1 cm]
2453778.14548 & 0.71 & 1.79\\[0.1 cm]
2453778.82453 & -26.82 & 1.71\\[0.1 cm]
2453837.86778 & 19.00 & 1.99\\[0.1 cm]
2453837.87490 & 17.74 & 1.83\\[0.1 cm]
2453841.86395 & 10.58 & 1.91\\[0.1 cm]
2453841.87031 & 8.42 & 1.79\\[0.1 cm]
2454130.10052 & -36.01 & 1.90\\[0.1 cm]
2454130.10671 & -37.32 & 1.90\\[0.1 cm]
2454130.99591 & 5.96 & 2.17\\[0.1 cm]
2454490.99221 & -31.71 & 1.93\\[0.1 cm]
2455197.95670 & 26.65 & 2.73\\[0.1 cm]
2455905.11046 & -31.07 & 2.00\\[0.1 cm]
2456641.07001 & -58.13 & 1.94\\[0.1 cm]
\noalign{\smallskip}
\hline
\end{tabular}
\tablefoot{\tablefoottext{*}{Barycentric dynamical time.}}
\end{table}

\clearpage
\section{Aliasing in HIRES data} \label{appendix:aliasinghires}
\begin{figure}
\centering
\includegraphics[width=1\columnwidth]{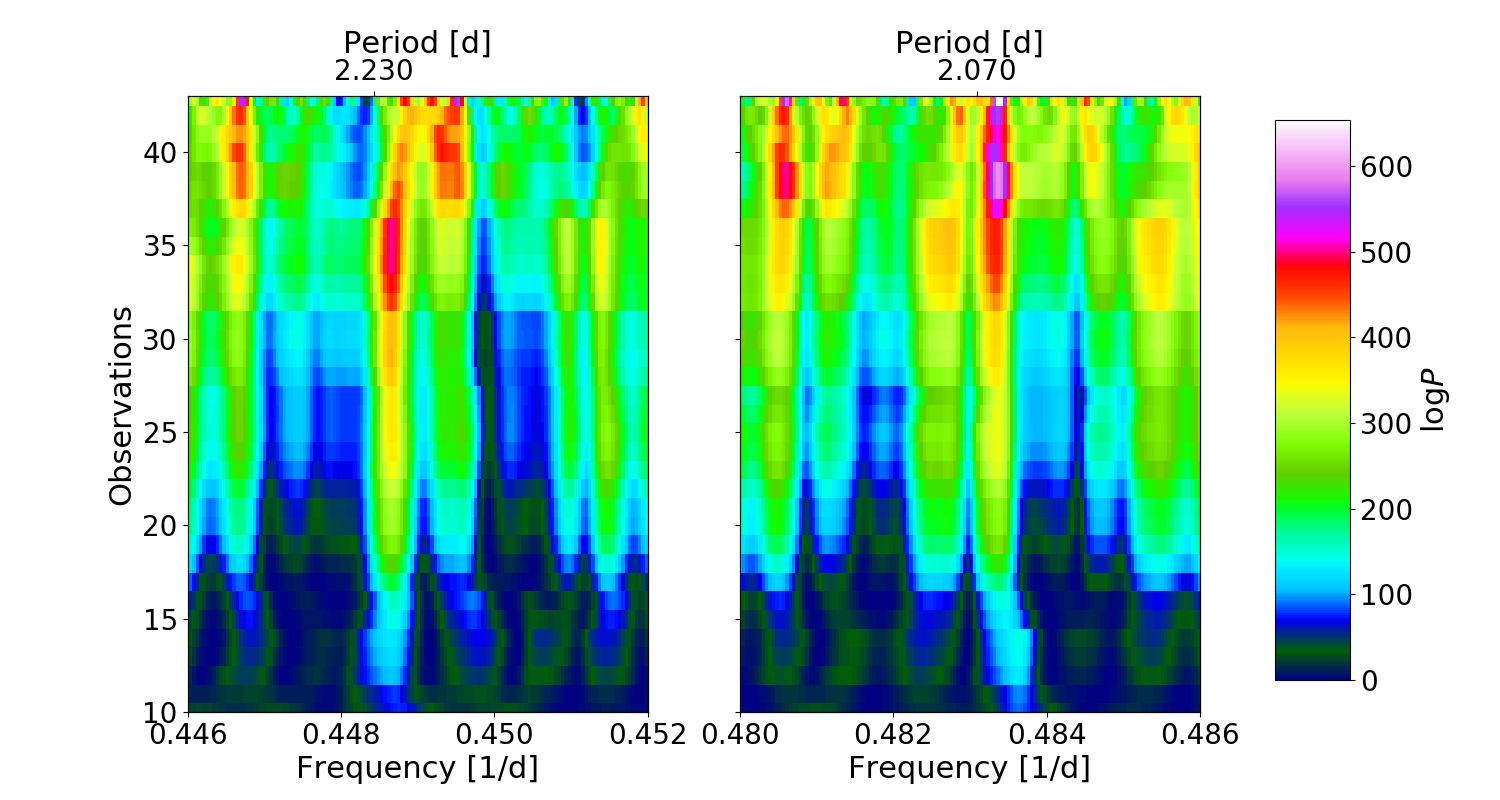}
\caption{S-BGLS periodograms of the HIRES data of AD~Leo, zoomed into two frequency ranges (left, 2.23\,d, and right, 2.07\,d) for the HIRES data. The nominal value $\log{P}$ does not carry significance, rather the relative values of $\log{P}$ are of importance.}
\label{fig:hiresbgls}
\end{figure}

The periodogram of the HIRES data set as a whole exhibits the strongest peak at 2.07\,d. 
instead of at 2.23\,d (see Fig.~\ref{fig:hiresbgls}). 
This can be understood as the window function has a peak at 1 month, more specifically, 29.53\,d; 2.07\,d and 2.23\,d are thus an alias pair with respect to this period.
As in \cite{DawsonFabrycky2010} and \cite{Stock2020_YZCeti}, a dealiasing approach using the \texttt{AliasFinder}\footnote{\url{https://github.com/JonasKemmer/AliasFinder}} \citep{Stock2020_aliasfinder} was performed. However, probably due to the fact that the underlying signal may not be a simple sinusoid, but rather a quasi-periodic signal with varying amplitudes and phaseshifts, the results were not fully conclusive. To further investigate, we performed a stacked Bayesian GLS (s-BGLS) periodogram (see Fig.~\ref{fig:hiresbgls}) where the stacking enables us to determine the coherence of a signal with increasing number of observations \citep{Mortier2015, Mortier2017}. As \cite{Mortier2017}, we normalized all s-BGLS periodograms to their respective minimum values, with the minimum probability set to 1. We found that until about 35 data points, which corresponds to the start of the sparse data, the 2.23\,d signal seems to be the most prominent. Afterwards, the alias at 2.07\,d gains significance. This behavior is in accordance to what we expect from an unstable, incoherent signal: when one signal loses significance, another one gains it \citep[as also demonstrated][with the Sun as an example]{Mortier2017}. We conclude that the HIRES data are fully consistent with an RV modulation with a 2.23\,d period.

 \clearpage
\section{Stellar activity}\label{sec:activityappendix}
 
\begin{figure*}[bp]
\centering
 
\includegraphics[width=.42\textwidth]{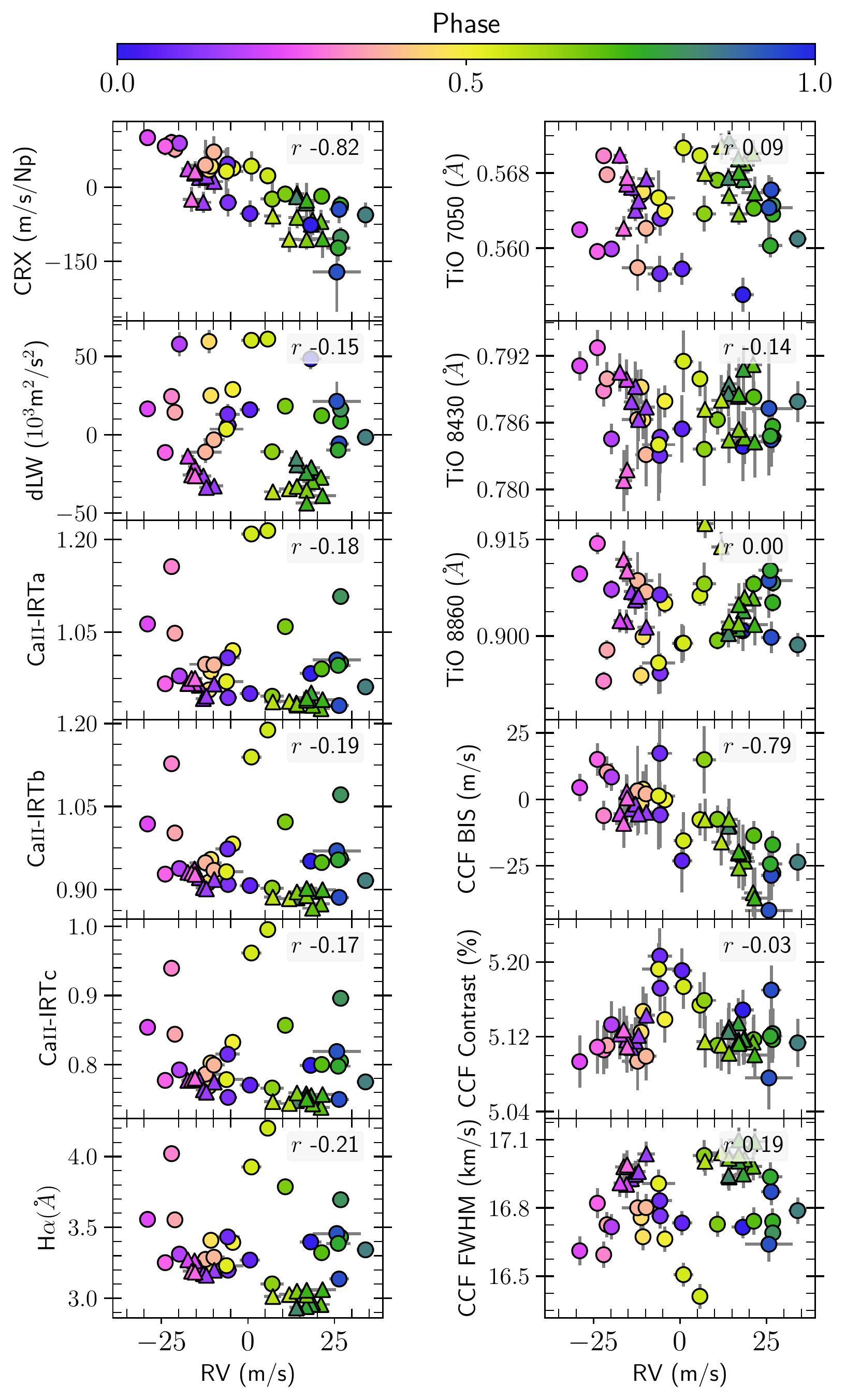}
\includegraphics[width=.42\textwidth]{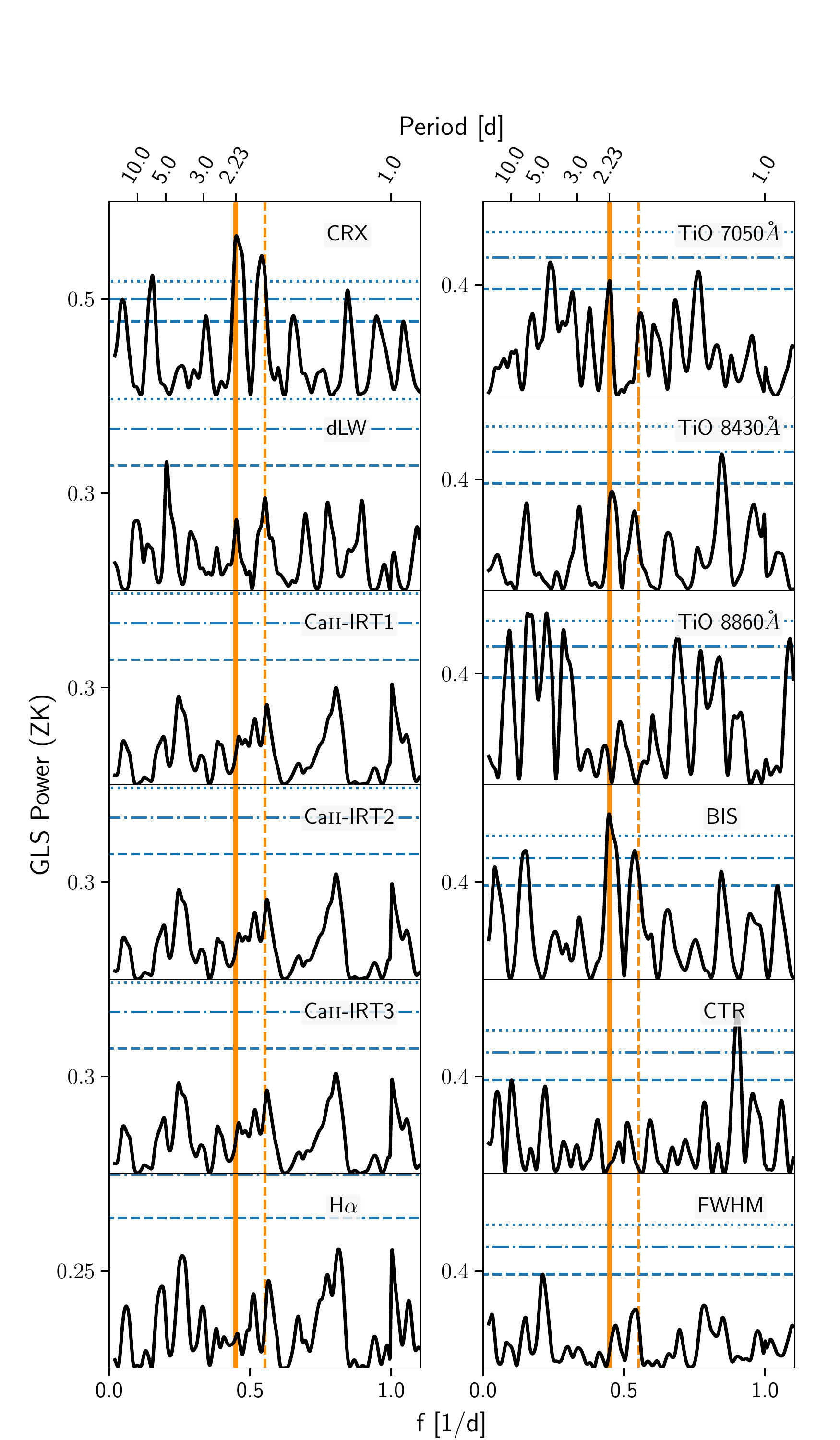}
 
\caption{Correlation plots with the RVs (\textit{left}) and GLS periodograms (\textit{right}) of the various stellar activity indicators from the CARMENES VIS spectroscopic data for AD~Leo. For the correlation plots, the circles and triangles represent the first and second subset of CARMENES VIS data, respectively. Data points are color-coded with the rotation phase. The Pearson-$r$ correlation coefficient combining both subsets is shown within each panel (R). For the periodograms, only the first season of the CARMENES VIS data was considered for plotting. The orange vertical solid and dashed lines represent the rotation period at $P=2.23$\,d and its daily alias at 1.81\,d. The horizontal dotted, dot-dashed, and dashed blue lines represent the 10\,\%, 1\,\%, and 0.1\,\% FAP levels.}
\label{fig:carmvisactivity}

\end{figure*} 

\begin{figure}
\centering
\includegraphics[width=.23\textwidth]{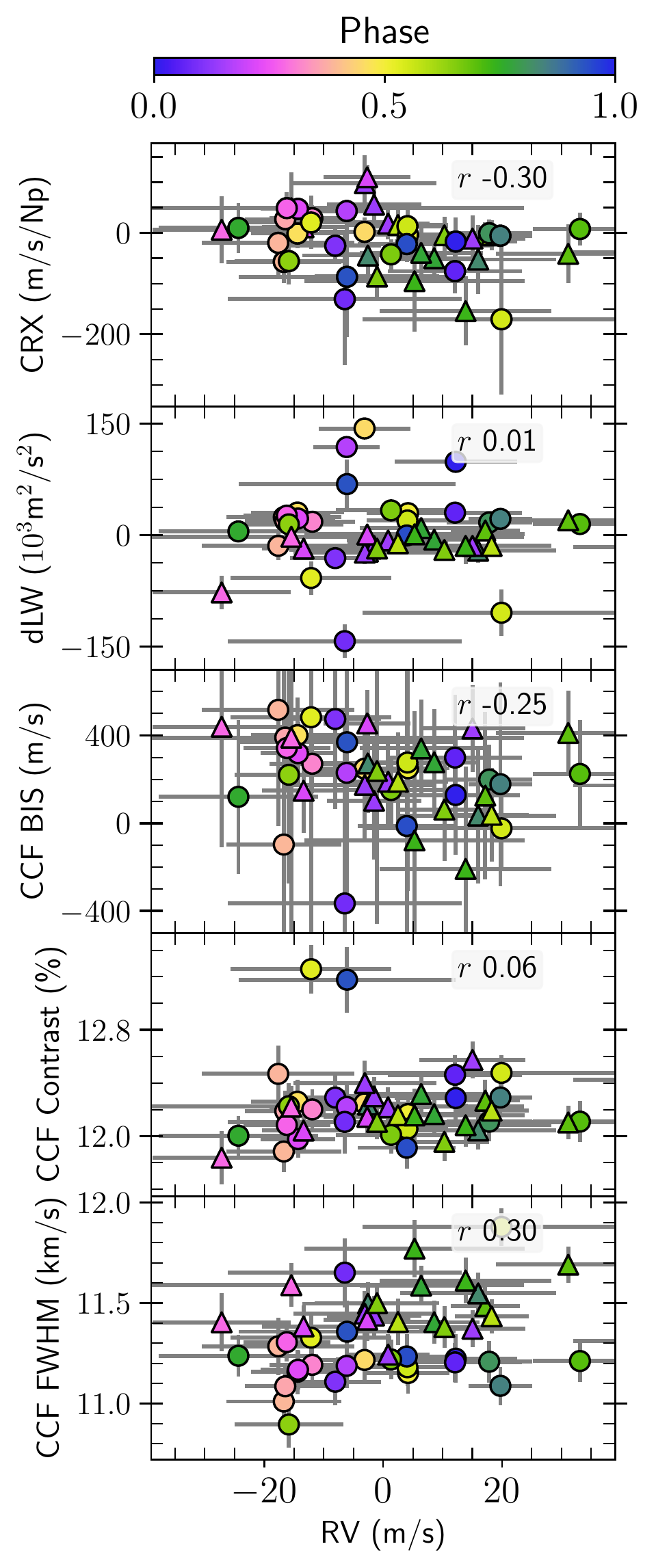}
\includegraphics[width=.23\textwidth]{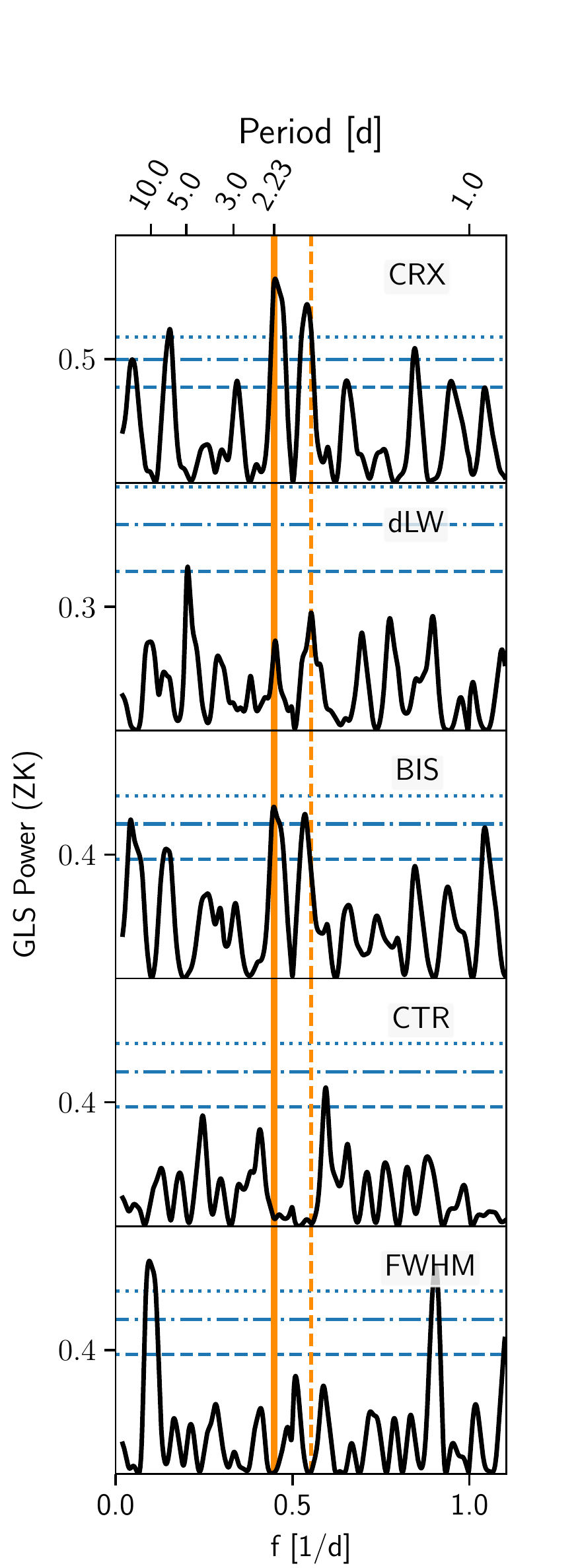}
\caption{Same as in Fig.~\ref{fig:carmvisactivity}, but for the CARMENES NIR channel.}
\label{fig:carmvniractivity}
\end{figure}

\begin{figure}
\centering
\includegraphics[width=.23\textwidth]{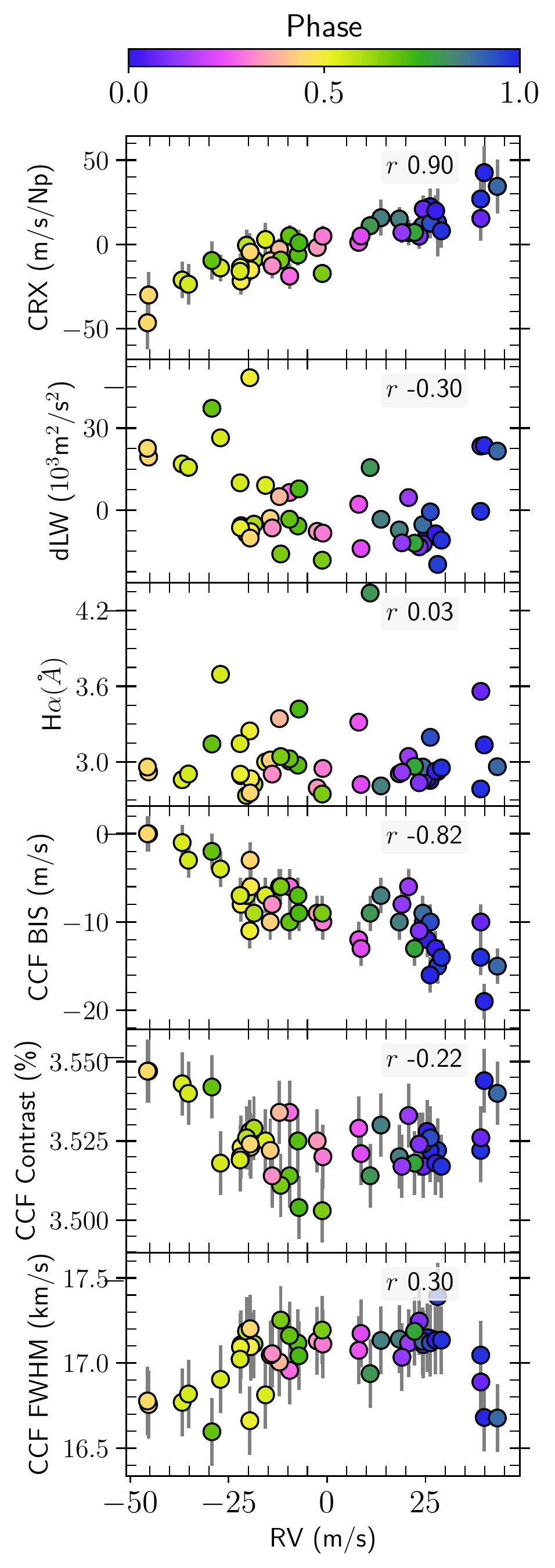}
\includegraphics[width=.23\textwidth]{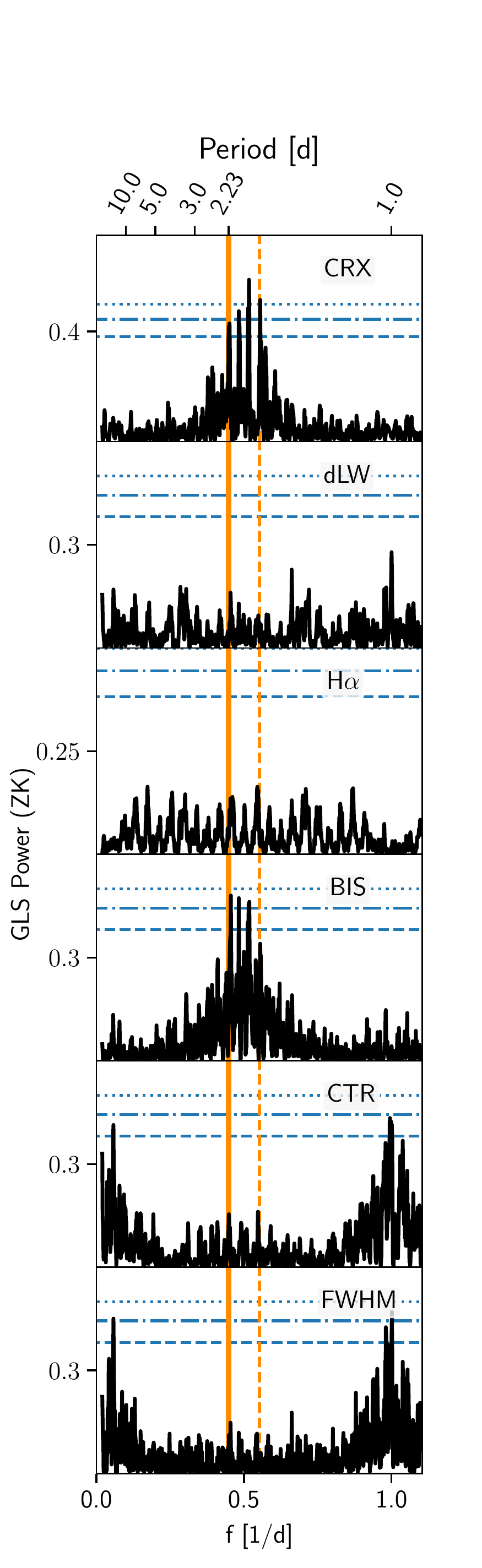}
\caption{Same as in Fig.~\ref{fig:carmvisactivity}, but for the HARPS instrument.}
\label{fig:harpsactivity}
\end{figure}

\end{appendix}

\end{document}